\documentclass[a4paper,11pt]{article}

\usepackage{youngtab}
\usepackage{graphicx,array}
\usepackage{color}
\usepackage{latexsym}
\usepackage{amsthm}
\usepackage{amsmath}
\usepackage{amssymb}
\usepackage{empheq}
\usepackage{mathtools}
\usepackage{scalerel,amssymb}

\usepackage{dsfont}
\usepackage{epsfig}
\usepackage{slashed}
\usepackage{bbold}
\usepackage{psfrag}
\usepackage[svgnames]{xcolor}
\PassOptionsToPackage{caption=false}{subfig}
\usepackage{subcaption}
\usepackage{xfrac}
\usepackage{multirow}
\usepackage{booktabs}
\usepackage{cite}
\usepackage[normalem]{ulem}
\usepackage{jheppub} 
\usepackage{colortbl}
\usepackage{array,booktabs,arydshln,xcolor}

\usepackage{youngtab}
\usepackage{graphicx,array}
\usepackage{color}
\usepackage{latexsym}
\usepackage{amsthm}
\usepackage{amsmath}
\usepackage{amssymb}
\usepackage{empheq}

\usepackage{dsfont}
\usepackage{epsfig}
\usepackage{slashed}
\usepackage{bbold}
\usepackage{psfrag}
\usepackage[svgnames]{xcolor}
\PassOptionsToPackage{caption=false}{subfig}
\usepackage{subcaption}
\usepackage{xfrac}
\usepackage{multirow}
\usepackage{booktabs}
\usepackage{cite}
\usepackage[normalem]{ulem}
\usepackage{jheppub} 

\newcommand{\be}{\begin{equation}}
\newcommand{\ee}{\end{equation}}
\newcommand{\beq}{\begin{equation}}
\newcommand{\eeq}{\end{equation}}
\newcommand{\bea}{\begin{eqnarray}}
\newcommand{\eea}{\end{eqnarray}}
\newcommand{\bei}{\begin{itemize}}
\newcommand{\eei}{\end{itemize}}

\newcommand{\lag}{{\cal L}}
\newcommand{\nn}{\nonumber}

\title{Perturbative unitarity constraints on generic vector interactions}

\author[a]{Daniele  Barducci}
\author[b]{Marco Nardecchia}
\author[b]{Claudio Toni }
\affiliation[a]{Universit\`a di Pisa and INFN Section of Pisa, Largo Bruno Pontecorvo 3, 56127, Pisa, Italia}
\affiliation[b]{Universit\`a degli Studi di Roma la Sapienza and INFN Section of Roma 1, Piazzale Aldo Moro 5, 00185, Roma, Italy}
\emailAdd{daniele.barducci@pi.infn.it}
\emailAdd{marco.nardecchia@roma1.infn.it}
\emailAdd{claudio.toni@uniroma1.it}

\abstract{
We study perturbative unitarity constraints on generic interactions between fermion and vector fields, which are allowed to have generic quantum numbers under a $\prod_i SU(N_i) \otimes U(1)$ group. We derive compact expressions for the bounds on the couplings for the cases where the fields transform under the trivial, fundamental or adjoint representation of the various, considering both the case of a complex vector arbitrary interactions with fermionic current and also the case of vectors arising as gauge fields. We apply our results to some specific NP models showing the constraints that can be derived using the tool of perturbative unitarity.
}

\begin{document} 
Last update \today
\maketitle

\section{Introduction and framework}\label{sec:intro}

Interactions between spin 1/2 matter fields and vector bosons, 
that we dub {\it vector interactions}, are one of the building blocks of renormalizable theories. For example they are a crucial ingredient of the Standard Model (SM) as they describe the interactions of quarks and leptons with the fields associated with the ${\cal G}_{\rm SM} = SU(3)_c \times SU(2)_L \times U(1)_Y$ gauge symmetry. While at the fundamental level these interactions arise from the local symmetry of the ultraviolet (UV) theory and the vector fields belong to the Adjoint representation of the local group, vector states can also emerge at low energy, without being gauge fields of a fundamental symmetry as in the case of QCD bound states. Vector interactions are also ubiquitous in New Physics (NP) theories that try to extend the SM in order to address its shortcomings,  see {\emph{e.g.}}~\cite{Langacker:2008yv} for a review.

New vector interactions are obviously constrained by many experimental searches, and the constraints are usually expressed as limits on a combination of couplings and masses and the bounds strongly depend on the experimental settings. On the other side the tool of perturbative unitarity (PU) allows to set an upper limit on the magnitude of the couplings, above which the perturbative expansion is expected to break down, independently on any experimental detail. In the past this technique, that we review in Sec.~\ref{sec:pu}, has been used to set an upper bound on the Higgs boson mass~\cite{Lee:1977yc,Lee:1977eg,Marciano:1989ns,Horejsi:2005da} 
and on the masses of quarks and leptons participating in weak interactions~\cite{Chanowitz:1978mv,Chanowitz:1978uj}\footnote{See also~\cite{Dicus:2004rg,Dicus:2005ku} for related works.} if weak  interactions were to remain weak at all energies. 
It has then been widely used in the literature to assess the range of validity of both renormalizable
 and effective operators~\cite{Griest:1989wd,Hally:2012pu,Kahlhoefer:2015bea,Chang:2019vez,Abu-Ajamieh:2020yqi,DiLuzio:2017tfn,DiLuzio:2016sur,Yin:2020afe,Capdevilla:2021rwo,Allwicher:2021jkr,DiLuzio:2017chi,Corbett:2014ora,Corbett:2017qgl,Almeida:2020ylr,Brivio:2021fog} or to derive useful sum rules on the couplings of theories~\cite{PhysRevD.43.904,Bento:2017eti,Bento:2018fmy,Bento:2023weq}.

In a recent work~\cite{Allwicher:2021rtd} we have used this tool to extract a bound on the couplings intervening in Yukawa type theories, where the field participating in the interaction had {\it arbitrary quantum number} under a ${\cal G}=\prod_i SU(N_i) \otimes U(1)$ group. We have build an explicit formalism and computed all the necessary ingredients for building the $2\to 2$ partial wave scattering matrix, namely the Lorentz parts of the scattering amplitudes and the group structure factors entering the amplitudes themselves, through which the PU limit could easily be derived.
In that work we have shown how the arising limits depend on both the dimensionality of the non abelian factors and on the representations of ${\cal G}$ under which the scalar and fermions fields transform. We have then applied our results to a set of  NP models formulated to explain some, at the time relevant, anomalous measurements in the magnetic moment of the muon $(g-2)_\mu$ and in semileptonic decays of $B-$meson both in neutral and in charged current interactions. 

In this paper we wish to extend our previous work~\cite{Allwicher:2021rtd} to the case of vector interactions and use the same formalism in order to answer the following question
\begin{quote}
 {\emph{Given an interaction between a vector and two fermions invariant under the  symmetry ${\cal G}=\prod_i SU(N_i) \otimes U(1)$, what is the maximum allowed value for the coupling with the requirement of PU? }}
\end{quote}
The choice to study $SU(N)$ is well motivated by its phenomenological application to gauge or flavour symmetries in SM or BSM theories. Differently from the case of Yukawa interactions studied in~\cite{Allwicher:2021rtd}, additional care is needed in order to study this scenario. In fact from one side in the high-energy limit some partial-waves present divergences that need to be regulated by appropriate, albeit arbitrary, choices of the kinematic region for the scattering processes, that will be discussed in the main part of the paper. From the other side if one consider the case of a vector field as a gauge field, trilinear and quartic self-interactions give additional contribution to the partial wave scattering matrix. We will deeply discuss these issues in the remaining of the paper, which is structured as follows. 
In Sec.~\ref{sec:pu} we briefly review the tool of PU and discuss the generic structure of vector interactions subject of our study.
Then in Sec~\ref{sec:eff_th} we examine the $2\to 2$ partial wave scatterings for theories were no self-interaction among the vector bosons is present and derive the corresponding PU limit, while in Sec.~\ref{sec:gauge} we discuss the case of vector gauge theories. We provide explicit calculation for a simple group $SU(N)$ that can be used as building block to generalise to larger groups such as $\prod_i SU(N_i) $, as shown in Sect. 5 of~\cite{Allwicher:2021rtd}. We then apply our results to two motivated NP theories in Sec.~\ref{sec:app} and then conclude in Sec.~\ref{sec:conc}. 
We also add few relevant appendices. In App.~\ref{app:wig} we report the explicit expressions of the Wigner $d-$functions, in App.~\ref{sec:LorPart} we report the Lorentz structure of the scattering amplitudes while in App.~\ref{App3} we report functions and group coefficients for computing PU bounds in the gauge case.

\section{General aspects}\label{sec:pu}

\subsection{The tool of perturbative unitarity}

For a $2\to 2$ scattering $i_1 i_2 \to f_1 f_2$ the partial waves, {\emph{i.e.}} the scattering amplitudes with fixed total angular momentum $J$, are defined as~\cite{Jacob:1959at}
\be\label{eq:partial_waves}
a_{fi}^J = 
\frac{
\beta_i^{1/4}(1, x_{i_1}, x_{i_2} )
\beta_f^{1/4}(1, x_{f_1}, x_{f_2} )}{32\pi}\int_{-1}^1 {\rm d}\!\cos\theta\, d_{\mu_i\mu_f}^J(\theta) {\cal T}_{fi}({\sqrt s, \cos\theta}) \ ,
\ee
where $\theta$ is the polar scattering angle in the center of mass frame and $\sqrt s$ the center of mass energy, $d_{\mu_i\mu_f}(\theta)$ are the small Wigner $d-$functions~\footnote{
Their explicit expressions are reported in App.~\ref{app:wig}
} where $\mu_i=\lambda_{i_1}-\lambda_{i_2}$ and $\mu_f=\lambda_{f_1}-\lambda_{f_2}$ are defined in terms of the helicities of the initial and final states, and $(2\pi)^4\delta^{(4)}(P_i -P_f)i {\cal T}_{fi}(\sqrt s, \cos\theta)=\langle f| S-1| i \rangle$, with $S$ the $S$-matrix. The kinematic factor $\beta_{i}(1, x_{i_1}, x_{i_2}) = 1 + x_{i_1}^2 + x_{i_2}^2 - 2 x_{i_1} - 2 x_{i_2} - 2 x_{i_1} x_{i_2}$ is a function of the initial state masses $x_{i_{1,2}} = m_{i_{1,2}}^2 / s $ and approaches unity in the high-energy massless limit, analogously to $\beta_f(1, x_{f_1}, x_{f_2})$,  giving for the partial wave
\be
a_{fi}^J = \frac{1}{32\pi}\int_{-1}^1 {\rm d}\cos\theta d_{\mu_i\mu_f}^J(\theta) {\cal T}_{fi}({\sqrt s, \cos\theta}) \ .
\ee
The unitarity fo the $S-$matrix implies the following relation among the partial waves
\be
\frac{1}{2i} (a_{fi}^J-a_{if}^{J*})=\sum_h a_{hf}^{J*}a_{hi}^J \ ,
\ee
where the sum runs over all the intermediate states $h$. By focusing on elastic channels $i=f$ and restricting the sum over $h$ only to $2-$particle states one obtains the condition
\be
{\rm Im}[a_{ii}^J] \ge |a_{ii}^J|^2 \ ,
\ee
which defines a circle in the complex plane, the Argand circle, inside which the amplitude must lie\footnote{Although this result is non perturbative, it may not hold in perturbation theory.}
\be
{\rm Re}^2[a_{ii}^J] +\left({\rm Im}[a_{ii}^J]-\frac{1}{2}\right)^2 \le \frac{1}{4} \ .
\ee
Since for $2\to 2$ high-energy forward scatterings the tree-level elastic amplitudes are real,  this suggests the following unitarity bound
\be\label{eq:pu-limit}
|a_{ii}^{J,{\rm tree}}| \le \frac{1}{2} \ .
\ee
where the arbitrary factor $1/2$ is a reasonable choice since a tree-level value which saturates Eq.~\eqref{eq:pu-limit} needs at least a higher-order correction of $\sim 40\%$ in order to re-enter the Argand circle, thus signalling the breakdown of the perturbative expansion. With this procedure, in order to extract the most stringent PU bound one needs to diagonalise the partial wave scattering matrix and identify the largest, in absolute value, eigenvalue, which correspond to the optimal elastic channel and that will set the PU limit.

\subsection{The structure of the vector interaction}\label{sec:structure_vec}

In this section we specify the class of models we are interested in and the assumptions we make in our analysis.
We take the mass terms for the fermion fields to be negligible and work in the  $\sqrt s \gg m_\psi$ limit, making massless Weyl fields as the natural degrees of freedom for the fermionic sector. On the other side we retain in our computation the masses of the vector fields. This is needed in order to avoid Coulomb singularities, which disappear when  one considers physical observables, but that are present in the certain tree-level expressions of the partial wave-amplitudes for massless vectors, as for the case of M\o ller and Bhabha scatterings in QED. Moreover this choice is also motivated by a theoretical beyond the SM (BSM) bias, in that we implicitly consider our results applicable to theories were the fermionic degrees of freedom are SM fields, while the vector fields are NP states, which generically have a larger mass than SM state\footnote{Clearly, this is not the only possibility, as light vector degrees of freedom are commonly present in SM extensions and subject of an intense theoretical and experimental investigation.}.
Under these assumptions we will express the partial wave amplitudes as function of the ratio between the boson mass $M$ and the center of mass energy, $x\equiv M^2/s$, and the derived constraint will in general depend on the value of $x$. As we will show, in order to not encounter Coulomb singularities we will need to set a lower limit on the value of the parameter $x$. In our analysis we adopt a pragmatic approach of considering $x \gtrsim 0.01$, given that the null results from present LHC searches enforce $M \gtrsim 1\;$TeV and that the next generation of collider searches, whether hadronic or leptonic, might happen at $\sqrt{s}\sim10\;$TeV.

Massive vector fields contain three physical polarization states. However in our analysis we will consider only the transverse  degrees of freedom, whose polarization vectors are independent of the mass, when vector fields are present as external initial and/or final state. Indeed the longitudinal modes
lead to amplitude that grow with the energy, which requires some sort of regularisation in the form of an analogous of the SM Higgs mode in a weakly coupled scenario or the exchange of heavier resonances in a strongly coupled one. However, in virtue of Cauchy interlace theorem\footnote{Cauchy interlace theorem  states that the largest eigenvalue of a reduced matrix is always $\le $ that the largest eigenvalue of the full matrix.}, by disregarding the longitudinal modes the bound that we will obtain will be conservative.

Moreover, in the case of $2\to 2$ scattering with $t-$ and/or $u-$channel contributions, singularities when the intermediate state can go on-shell can appear. This happens if $\sqrt s$ is within a certain interval fixed by the scattering masses. We can think at these amplitudes as a sequence of two $1\to 2$ and $2\to 1$ subprocesses~\cite{Grzadkowski:2021kgi}. These singularities give rise to divergent cross-sections.
There is no unique and well defined procedure to follow in order to regulate these singularities.
In some cases, it is enough to add a finite width for the propagating particle, in other cases it is necessary to consider the specific physics case and the {\it environment} of the system under examination. For example in \cite{Melnikov:1996na,Melnikov:1996iu}, the finite beams size makes the physical cross sections finite, while in 
cosmological applications, the singularity can be regulated by the properties of the thermal bath \cite{Grzadkowski:2021kgi}. 
Our approach will be to avoid these singularities by restricting the possible range of values of the parameter $x$. As we will show, in our case such prescription is required only in the $\psi V \to \psi V$ scattering channel, where we will need to only consider $x>1/2$.

The interaction that we consider in this work are described by the sum of two terms: a coupling between the fermion and vector fields $\lag_{\psi}$ and, possibly, the self-interactions of the vector fields $\lag_{\text{self}}$, {\emph{i.e.}}
\be
\lag=\lag_{\psi}+\lag_{\text{self}} \ , 
\ee
where in all generality we can write, in analogy with~\cite{Allwicher:2021rtd},
\be\label{eq:gen}
\lag_{\psi} = {\cal G}_{\alpha i j} V_\mu^\alpha \bar \psi_{L,i} \gamma^\mu \psi_{L,j} \ , \qquad {\cal G}_{\alpha i j} = {\cal G}^*_{\alpha j i} \ ,
\ee
where $V_\mu$ is a set of $N_V$ real vector fields and $\psi_L^i$ a set of $N_\psi$ left-handed Weyl fermions fields. The ${\cal G}_\alpha$ matrix is hermitian in the real vector basis. The hermiticity condition ${\cal G}_{\alpha i j} = {\cal G}^*_{\alpha j i}$ implies that no explicit $h.c.$ is needed in Eq.~\eqref{eq:gen}.
Interactions between right-handed fields can be written in this formalism by using the charge-conjugation operator. For example if $\psi_L = \eta_R^c$ one has that $\bar \eta_R^c \gamma_\mu \eta_R^c = - \bar \eta_R \gamma_\mu \eta_R$. 

In order to understand the structure of the $2\to 2$ tree-level scattering matrix, lets consider it in the $\{\psi\psi, \psi V, VV \}$ basis where it can be schematically written as
\vskip 10pt
\be\label{eq:scatt_mat_gen} {\cal T}_{fi} = 
     \left(
    \begin{array}{c !{\vrule width 1pt}  c | c | c}
           &   \psi\psi  & \psi V    &VV       \\
 \specialrule{1pt}{0pt}{0pt}    
\psi\psi         &  	 {\cal T}_{\psi\psi \to \psi \psi}		 &   				& 	{\cal T}_{VV \to \psi \psi}		    \\
\psi V 	   &    		& 		 {\cal T}_{\psi V \to \psi V}	 	&		  \\
V V 		   &	 {\cal T}_{\psi \psi \to VV}  		& 				 &	 {\cal T}_{V V \to VV}   		   \\
    \end{array}
    \right) \ .
\ee
\vskip 10pt
Here the white blocks are zero because of Lorentz invariance, while the others receive contribution from both the interaction of Eq.~\eqref{eq:gen} and also, possibly, from ${\cal L}_{\rm self}$\footnote{The topologies present only in the case of trilinear and quartic vector interactions contained in $\lag_{\text{self}}$ are similar to the one that would have been present by also  considering  a scalar potential in~\cite{Allwicher:2021rtd}. Moreover they depend on the UV nature of the vector field.}. More precisely ${\cal T}_{\psi\psi \to \psi \psi}$ receives only contribution from the interaction of Eq.~\eqref{eq:gen} and is always kinematically allowed, while ${\cal T}_{\psi\psi \to V V}$ receives contribution both from Eq.~\eqref{eq:gen} but also from Feynman diagram with a trilinear $V_\mu$ interaction with a vector in $s-$channel and is kinematically allowed for $x<1/4$. Furthermore ${\cal T}_{\psi V \to \psi V}$ receives contribution both from Eq.~\eqref{eq:gen} but also from Feynman diagram with a trilinear $V_\mu$ interaction with a vector in $t-$channel and is kinematically allowed for $x<1$ and finally ${\cal T}_{V V \to V V}$ receives contribution only from Feynman diagram with a trilinear and quartic $V_\mu$ interactions and is kinematically allowed for $x<1/4$.

As in~\cite{Allwicher:2021rtd} we will derive the group structure factors entering the amplitudes by assuming the fields to be charged under a {\it single} $SU(N)$ factor, which will serve as building blocks for more complicated theories, as explicitly shown in~\cite{Allwicher:2021rtd}. In the following we will distinguish between two different scenarios
\begin{itemize}
\item We first consider the case a vector field transforming under a complex representation of the $SU(N) \otimes U(1)$ group and only focus  on the interaction given by Eq.~\eqref{eq:gen}. Since a complex vector field with $U(1)$ charge has no trilinear self coupling, we can 
restrict our study to the 
${\cal T}_{\psi\psi \to \psi \psi}$ and ${\cal T}_{\psi V \to \psi V}$  channels of the scattering matrix of Eq.~\eqref{eq:scatt_mat_gen}, where the former has integer $J$, while the latter semi-integer $J$.
Lorentz invariance implies that these two scatterings do not mix among themselves, while
Cauchy interlace theorem guarantees that by neglecting the $\psi \psi \to VV$ and $V V\to VV$ channels one obtains a conservative bound. This simplifies the analysis.
\item We then consider the case of an $SU(N)\otimes U(1)$ gauge vector field, whose self-interaction $\lag_{\text{self}}$ is completely fixed by the gauge symmetry and consider the scattering matrix of Eq.~\eqref{eq:scatt_mat_gen} in full generality.
\end{itemize}

As anticipated, the analysis drastically simplifies by decomposing the $2\to2$ scattering amplitude into a Lorentz part which depends only on the spin and helicity of the involved fields and a group-theoretical part that depends on the $SU(N)$ quantum number~\cite{Allwicher:2021rtd}. Additionally, the $U(1)$ symmetry enforces useful selection rules in the case of a complex vector field. More practically any $2\to2$ scattering amplitude among particles $i_{1,2}$ and  $f_{1,2}$ can be decomposed as
\be\label{eq:total_T}
{\cal T}_{f_1 f_2 i_1 i_2}^{\lambda_{f_1}\lambda_{f_2}\lambda_{i_1}\lambda_{i_2}} (\sqrt s, \theta)=\bigoplus_{{\bf{r}}} \sum_{m=s,t,u} {\cal T}^{\lambda_{f_1}\lambda_{f_2}\lambda_{i_1}\lambda_{i_2}}_m(\sqrt s,\theta) {\cal F}^{m, {\rm {\bf{r}}}}_{f_1 f_2 i_1 i_2} (N) {\mathbb 1}_{d_{{\bf{r}} }}\, ,
\ee
where ${\cal T}^{\lambda_{f_1}\lambda_{f_2}\lambda_{i_1}\lambda_{i_2}}_m(\sqrt s,\theta)$ is the Lorentz part of the scattering amplitude and ${\cal F}^{m, {\rm {\bf{r}}}}_{f_1 f_2 i_1 i_2} (N)$ is the group factor that include\footnote{We also include into the group factor a $1/\sqrt{2}$ factor for each (final or initial) state composed by identical particles.} any information of the $SU(N)$ $\bf{r}$ irreducible representation, while $d_{{\bf{r}} }$ stands for the dimensionality of $\bf{r}$. The Lorentz part of the scattering amplitudes needed for deriving the PU bounds can be found in Appendix~\ref{sec:LorPart}.

\section{Complex vector theories}\label{sec:eff_th}

In this Section we consider models where the vector fields are in complex representation of the $SU(N)\otimes U(1)$ group with non vanishing $U(1)$ charge.
This means that we need at least two type of fermion field to write an interaction, which reads\footnote{The parameter $g$ is generally complex but we can absorb its phase redefining one of the fields and thus set $g$ to be real and non negative without loss of generality.}
\be\label{eq:complex}
{\cal L}_{\rm complex} = g\bar\chi\gamma^\mu\xi V_\mu + g\bar\xi\gamma^\mu\chi V_\mu^\dag \ ,
\ee
where, without loss of generality, $\chi$ and $\xi$ are left-handed fermion fields and, differently from Eq.~\eqref{eq:gen}, we have written the interaction in the complex vector basis, 
since the presence of the $U(1)$ symmetry allows to simplify the scattering structure, thanks to selection rules that forbids certain scattering channels. As already stressed, the complex nature of the  vector field forbids a trilinear self coupling, while quartic vector interactions are still possible, and hence $V V\to VV$ scattering can in principle be present.
For simplicity however in the case of integer $J$ we only consider the scattering sub-matrix $\cal T_{\psi\psi \to \psi \psi}$ that is controlled by Eq.~\eqref{eq:complex}, in virtue of Cauchy interlace theorem that guarantees to obtain a conservative bound.
We now first review the general properties of the two scattering matrices,  without specifying the quantum numbers of the involved fields. Then we will consider explicit models with different quantum number assignments and show the constraint from PU requirement for each one.

\subsection{Integer $J$ scattering: $\psi\psi \to \psi \psi$}\label{sec_integer_scatt}

\begin{figure}
	\centering
	\includegraphics[scale=0.8]{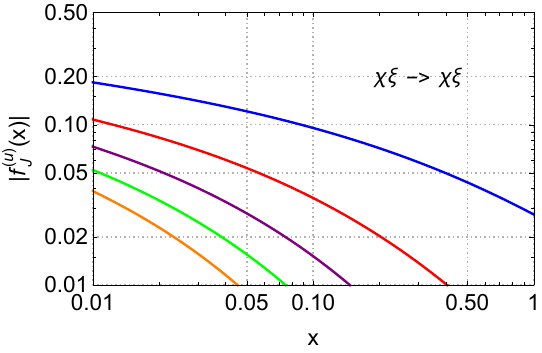}
	\hfill
	\includegraphics[scale=0.8]{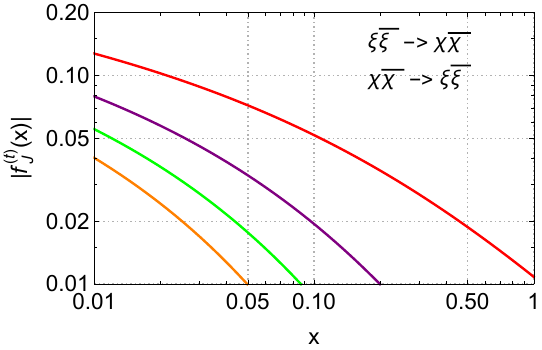}
	\caption{ $|f^{(u)}_{J}(x)|$ (left) and $|f^{(t)}_{J}(x)|$ (right) functions defined in Eq.~\eqref{eq:complex_4psi_f} for $J=0$ (blue), $J=1$ (red), $J=2$ (purple), $J=3$ (green), $J=4$ (orange).}
	\label{fig:fut}
\end{figure}

The scattering matrix form in the helicity basis reads
\vskip 10pt
\be\label{eq:full_matrix}
    {\cal T}_{\psi\psi\to\psi\psi} =
     \left(
    \begin{array}{c c !{\vrule width 1pt} c c|c}
           &     &   \multicolumn{2}{c|}{\mu_i=0} &  \mu_i=-1   \\
         &  &  ++ 				& -- & -+  \\
\specialrule{1pt}{0pt}{0pt}
\multirow{2}{*}{$\mu_f=0$} & ++   & {\cal T}^{++++}  &   &      \\
 &--   &  	 &  {\bf {\cal T}}^{----}  &      \\
\hline
\mu_f=-1 &-+   &  	 &  &  {\bf {\cal T}}^{-+-+}      \\
    \end{array}
    \right) \ ,
\ee
\vskip 10pt
\noindent where the Lorentz part of the non zero scatterings are collected in App.~\ref{sec:LorPart}.
By using Eq.~\eqref{eq:total_T}, the non vanishing partial wave scattering matrices in a given representation ${\bf r}$ are given by
\begin{align}
& a_{\chi\xi\chi\xi}^{J,\mathbf{r}}  =g^2 {\cal F}^{u, {\rm {\bf{r}}}}_{\chi\xi\chi\xi} f^{(u)}_{J}(x) \, , 
\nn \\
& a_{\chi\bar\xi\chi\bar\xi}^{J,\mathbf{r}}=g^2 {\cal F}^{s, {\rm {\bf{r}}}}_{\chi\bar\xi\chi\bar\xi} f^{(s)}_{J}(x,\gamma) \, ,  \nn \\
& a_{\xi\bar\chi\xi\bar\chi}^{J,\mathbf{r}}=g^2 {\cal F}^{s, {\rm {\bf{r}}}}_{\xi\bar\chi\xi\bar\chi} f^{(s)}_{J}(x,\gamma)  \nn \\
& a_{\{\chi\bar\chi,\xi\bar\xi\}\{\chi\bar\chi,\xi\bar\xi\}}^{J,\mathbf{r}}  =g^2 f^{(t)}_{J}(x) \begin{pmatrix} 0 & {\cal F}^{t, {\rm {\bf{r}}}}_{\chi\bar\chi\xi\bar\xi} \\ {\cal F}^{t, {\rm {\bf{r}}}}_{\xi\bar\xi\chi\bar\chi} & 0 \end{pmatrix} \, ,
\end{align}
where the subscript indicates the chosen basis, $J$ is an integer and the functions here introduced are given by\footnote{$P_{n}^{(\alpha,\beta)}(z)$, with $n$ being a non negative integer, are the Jacobi polynomials. For negative $n$, read $P_{n}^{(\alpha,\beta)}(z)=0$.}
\begin{align}\label{eq:complex_4psi_f}
f^{(s)}_{J}(x,\gamma)=& -\frac{1}{24\pi}\frac{\delta_{J1}}{1-x+i\gamma x} \, , \nn \\
f^{(u)}_{J}(x)=& -\frac{1}{8\pi}\int_{-1}^{1}\!\text{d}{z}\, P_{J}^{(0,0)}(z)\frac{1}{1+z+2x} \, , \nn \\
f^{(t)}_{J}(x)=& \frac{1}{32\pi}\int_{-1}^{1}\!\text{d}{z}\, P_{J-1}^{(0,2)}(z)\frac{(1+z)^2}{1-z+2x} \, ,
\end{align}
with $\gamma\equiv\Gamma/M$ the ratio between the boson decay width and its mass. The benchmark value for this parameter from the decay width in the massless fermion limit is $\gamma=\Gamma(V\to\chi\bar\xi)/M=g^2/(12\pi)$, so that the $s$-channel contribution is independent on the coupling $g$ near the resonance, {\emph{i.e.}} for $x\simeq1$.
Note that the selection rules imposed by the $U(1)$ symmetry, or equivalently the complex nature of the vector field, separate the contributions of the $s-$, $t-$ and $u-$channels. Each scattering receives contribution from only one of these channels and the group factor is then factorised. In Fig.~\ref{fig:fut} we show the values of $|f^{(u)}_{J}(x)|$ and $|f^{(t)}_{J}(x)|$ for different values of $J$. Note that the scatterings with $\mu_i=\mu_f=-1$ have null projection on the $J=0$ partial amplitude.
Recalling that we have chosen to focus on $x>0.01$ as a reasonable choice to avoid Coulomb singularities, we can extract the strongest constraints on $g$ by maximising the value of the partial wave scattering amplitude eigenvalues over the parameters $x$ and $J$. By a numerical computation one obtains
\be
g\lesssim \text{Min}\left\{ \frac{1.65}{\sqrt{|{\cal F}^{u, {\rm {\bf{r}}}}_{\chi\xi\chi\xi}|}},\frac{\sqrt{12\pi}}{\sqrt{|{\cal F}^{s, {\rm {\bf{r}}}}_{\chi\bar\xi\chi\bar\xi}|}},\frac{1.98}{\sqrt{|{\cal F}^{t, {\rm {\bf{r}}}}_{\chi\bar\chi\xi\bar\xi}|}} \right\} \, ,
\ee
which depends on the group factors of the models, that will be computed in Sec.~\ref{sec:complex_group}. As it can be seen from the figure it is the lowest value of $J$ that enables to set the strongest PU bound, while this can arise from different channels, depending on the value of $N$.

\subsection{Half-integer $J$ scattering: $\psi V \to\psi V$}\label{sec_half_integer_scatt}

\begin{figure}
	\centering
	\includegraphics[scale=0.8]{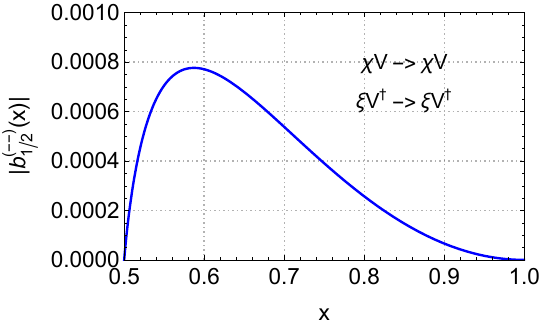}
	\hfill
	\includegraphics[scale=0.8]{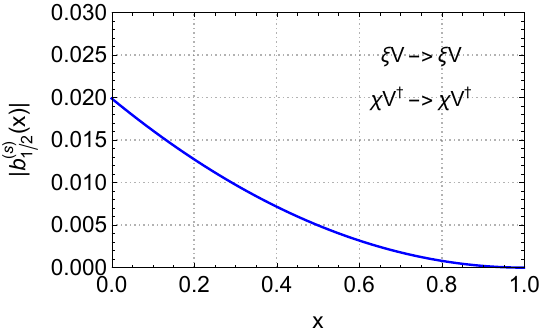}
	\caption{
	$b^{(--)}_{1/2}(x)$  (left) and $b^{(s)}_{1/2}(x)$ (right) functions defined in Eq.~\eqref{eq_b_func}.}
	\label{fig:bsmm}
\end{figure}

The scattering matrix form in the helicity basis now reads
\vskip 10pt
\be\label{bho}
    {\cal T}_{\psi V \to\psi V} =
     \left(
    \begin{array}{c c !{\vrule width 1pt} c|c|c|c}
           &     &   \mu_i=\frac{3}{2} &   \mu_i=\frac{1}{2} &   \mu_i=-\frac{1}{2} &   \mu_i=-\frac{3}{2}   \\
           &     &  +-    & --      & ++ & -+  \\
\specialrule{1pt}{0pt}{0pt}
\mu_f=\frac{3}{2} & +-   & {\cal T}^{+-+-}	 &  &  {\cal T}^{+-++} &   \\
\hline
\mu_f=\frac{1}{2} & --   & 	 & {\cal T}^{----}  &   &  {\cal T}^{---+} \\
\hline
\mu_f=-\frac{1}{2} & ++   & {\cal T}^{+++-}	 &   & {\cal T}^{++++}  &   \\
\hline
\mu_f=-\frac{3}{2} & -+   & 	 &  {\cal T}^{-+--} &   & {\cal T}^{-+-+}  \\
    \end{array}
    \right) \ .
\ee
\vskip 10pt
Note that  in this case the kinematic\footnote{Recall that $s>(m_{i_1}+m_{i_2})^2, (m_{f_1}+m_{f_2})^2$.} fixes $x<1$.
Since we are assuming no vector self-interaction, the scattering only proceed via $s-$ and $u-$channel topologies, for which we report the Lorentz parts in App.~\ref{sec:LorPart}. The non vanishing partial scattering matrices in a given representation $\mathbf{r}$ are now given by
\begin{align} \label{eq:apvpvb}
& a_{\xi V_{-1}\xi V_{-1}}^{J,\mathbf{r}}=g^ 2 {\cal F}^{s, {\rm {\bf{r}}}}_{\xi V\xi V} b^{(s)}_{J}(x) \,  \nn \\
& a_{\chi V_{-1}^\dag\chi V_{-1}^\dag}^{J,\mathbf{r}}=g^ 2 {\cal F}^{s, {\rm {\bf{r}}}}_{\chi V^\dag\chi V^\dag} b^{(s)}_{J}(x) \, , \nn \\
& a_{\{\xi V_{-1}^\dag,\xi V_{+1}^\dag\}\{\xi V_{-1}^\dag,\xi V_{+1}^\dag\}}^{J,\mathbf{r}}=g^2 {\cal F}^{u, {\rm {\bf{r}}}}_{\xi V^\dag\xi V^\dag} \begin{pmatrix} b^{(--)}_{J}(x) & b^{(-+)}_{J}(x) \\ b^{(-+)}_{J}(x) & b^{(++)}_{J}(x) \end{pmatrix} \, ,\nn  \\
& a_{\{\chi V_{-1},\chi V_{+1}\}\{\chi V_{-1},\chi V_{+1}\}}^{J,\mathbf{r}}=g^2 {\cal F}^{u, {\rm {\bf{r}}}}_{\chi V\chi V} \begin{pmatrix} b^{(--)}_{J}(x) & b^{(-+)}_{J}(x) \\ b^{(-+)}_{J}(x) & b^{(++)}_{J}(x) \end{pmatrix} \, ,
\end{align}
where the functions here introduced are given by
\begin{align}\label{eq_b_func}
& b^{(s)}_{J}(x)= -\frac{\delta_{J\frac{1}{2}}}{16\pi}(1-x)^{2} \, , \nn \\
& b^{(--)}_{J}(x)= \frac{1}{32\pi}\int_{-1}^{1}\!\text{d}{z}\, P_{J-\frac{1}{2}}^{(0,1)}(z)\frac{(1-x)^{2}(1-2x)(1-z^2)}{2x^2-(1-x)^2(1+z)} \, , \nn \\
& b^{(-+)}_{J}(x)= \frac{1}{64\pi}\sqrt{\frac{2J+3}{2J-1}}\int_{-1}^{1}\!\text{d}{z}\, P_{J-\frac{3}{2}}^{(2,1)}(z)\frac{x(1-x)^{2}(1+z)(1-z)^2}{2x^2-(1-x)^2(1+z)} \, , \nn \\
&
b^{(++)}_{J}(x)= \frac{1}{64\pi}\int_{-1}^{1}\!\text{d}{z}\, P_{J-\frac{3}{2}}^{(0,3)}(z)\frac{(1-x)^{2}(1+z)^3}{2x^2-(1-x)^2(1+z)} \, .
\end{align}
Except than for $b^{(s)}_{J}(s)$, all these functions are well defined only in the region $x>1/2$ since the intermediate state can otherwise go on-shell, thus giving rise to a singularity for $b^{(-+)}_{1/2}(x)$ and $b^{(++)}_{1/2}(x)$ for $J>1/2$. We then choose to restrict our analysis to the 
 $J=1/2$ amplitudes, for which the $b^{(-+)}_{1/2}(x)$ and $b^{(++)}_{1/2}(x)$ functions are absent, making the $2\times 2$ scattering matrices defined above trivially diagonal. We show the behaviour of the $|b^{(s)}_{1/2}(x)|$ and $|b^{(--)}_{1/2}(x)|$ function in Fig. \ref{fig:bsmm}. By taking the maximal value of these function in the allowed region, we can extract the bounds
\be
g\lesssim \text{Min}\left\{ \frac{25.38}{\sqrt{|{\cal F}^{u, {\rm {\bf{r}}}}_{\xi V^\dag\xi V^\dag}|}}, \ \frac{25.38}{\sqrt{|{\cal F}^{u, {\rm {\bf{r}}}}_{\chi V\chi V}|}}, \ \frac{\sqrt{8\pi}}{\sqrt{|{\cal F}^{s, {\rm {\bf{r}}}}_{\xi V\xi V}|}}, \ \frac{\sqrt{8\pi}}{\sqrt{|{\cal F}^{s, {\rm {\bf{r}}}}_{\chi V^\dag\chi V^\dag}|}} \right\} \, ,
\ee
which again depends on the group factors of the models that will be computed in Sec.~\ref{sec:complex_group}.

\subsection{Group factors}\label{sec:complex_group}

In this section we compute the group factors ${\cal F}^{m, {\rm {\bf{r}}}}_{f_1 f_2 i_1 i_2} (N) $ for different quantum number assignments of the various fields.
For concreteness, we consider only the case of singlet, fundamental and Adjoint $SU(N)$ representations, further restricting the models to the cases of a single adjoint. The calculation of the group factors follows the same procedure of~\cite{Allwicher:2021rtd}.

\subsubsection{First model: $\chi\sim {\tiny{\fbox{$\protect\phantom s$} }}_q$, $\xi \sim {\tiny{\fbox{$\protect\phantom s$} }}_{q^\prime}$, $V\sim \mathbf{1}_{q-q^\prime}$ }\label{sec:model1}

In this model with $q\ne q^\prime$, $V$ is a complex $SU(N)$ singlet while $\chi$ and $\xi$ transform under the fundamental representation of the $SU(N)$ group. The fermionic two particle states decompose as
\begin{align}\label{eq:model1_dec}
& \chi\xi\sim{\rm {\bf{S}}}+{\rm {\bf{AS}}}\, , \nn\\
& \chi\bar\xi\sim\chi\bar\chi\sim\xi\bar\xi\sim\mathbf{1}+\mathbf{Adj}\, ,
\end{align}
where we indicate with ${\rm {\bf{S}}}$ and {\rm {\bf{AS}}} the totally symmetric and antisymmetry irreducible representations that arise from tensor decomposition of the product ${\tiny{\yng(1)}}\times{\tiny{\yng(1)}}$\ .
Other two particle states trivially decompose since $V$ is a singlet.
The group factors for this model are reported in Tab.~\ref{GF1}. Following the results of
Sec.~\ref{sec_integer_scatt} and Sec.~\ref{sec_half_integer_scatt}
the strongest constraint on the coupling $g$ in this model is then
\be
g\lesssim\text{Min}\left\{ 1.65, \ 
\sqrt{\frac{12\pi}{N}}
\right\} \, .
\ee

\begin{table}[t!]
\begin{center}
\begin{tabular}{cc||c|c|c|c|c|c|c}
\multicolumn{2}{c||}{{\bf First model}} & ${\cal F}^{m, \mathbf{r}}_{\chi\xi\chi\xi}$ & ${\cal F}^{m, \mathbf{r}}_{\chi\bar\xi\chi\bar\xi}$ & ${\cal F}^{m, \mathbf{r}}_{\chi\bar\chi\xi\bar\xi}$ & ${\cal F}^{m, \mathbf{r}}_{\xi V\xi V}$ & ${\cal F}^{m, \mathbf{r}}_{\chi V\chi V}$ & ${\cal F}^{m, \mathbf{r}}_{\xi V^\dag\xi V^\dag}$ & ${\cal F}^{m, \mathbf{r}}_{\chi V^\dag\chi V^\dag}$ \\
\midrule \midrule
$m=s$ & $\mathbf{r}=\mathbf{1}$ &  & $N$ & & & & &  \\
\midrule
 & $\mathbf{r}={\tiny{\yng(1)}}$ &  &  & & 1 &  & & 1 \\
\hline \hline
$m=t$ & $\mathbf{r}=\mathbf{1}$ &  &  & 1 & & & &   \\
\hline
 & $\mathbf{r}=\mathbf{Adj}$ &  &  & 1 & & & &   \\
\hline\hline
$m=u$ & $\mathbf{r}={\rm {\bf{S}}}$ & 1 &  &  & & & &   \\
\hline
 & $\mathbf{r}={\rm {\bf{AS}}}$ & $-1$ &  & & & & &    \\
\hline
 & $\mathbf{r}={\tiny{\yng(1)}}$ & &  & & & 1 & 1 &    \\
\end{tabular}
\end{center}
\caption{Group factors coefficient for the first model of the complex vector case. Here $\chi\sim {\tiny{\fbox{$\protect\phantom s$} }}_q$, $\xi\sim {\tiny{\fbox{$\protect\phantom s$} }}_{q^\prime}$, $V\sim \mathbf{1}_{q-q^\prime}$.}
\label{GF1}
\end{table}

\subsubsection{Second model: $\chi\sim {\tiny{\fbox{$\protect\phantom s$} }}_q$, $\xi \sim \mathbf{1}_{q-q^\prime}$, $V \sim {\tiny{\fbox{$\protect\phantom s$} }}_{q^\prime}$}\label{sec:model2}

In this model $V$ is always a complex field, regardless of the values of $q-q^\prime$. The two particle states decompose similarly to Eq.~\eqref{eq:model1_dec} and the group factors are reported in Tab.~\ref{GF2}. The strongest constraint on the coupling $g$ turns then out to be
\be
g\lesssim\text{Min}\left\{ 1.65, \ \frac{1.98}{N^{1/4}}, \ 
\sqrt{\frac{8\pi}{N}}
\right\} \, .
\ee

\begin{table}[t!]
\begin{center}
\begin{tabular}{cc || c|c|c| c|c|c|c}
\multicolumn{2}{c||}{{\bf Second model}} & ${\cal F}^{m, \mathbf{r}}_{\chi\xi\chi\xi}$ & ${\cal F}^{m, \mathbf{r}}_{\chi\bar\xi\chi\bar\xi}$ & ${\cal F}^{m, \mathbf{r}}_{\chi\bar\chi\xi\bar\xi}$ & ${\cal F}^{m, \mathbf{r}}_{\xi V\xi V}$ & ${\cal F}^{m, \mathbf{r}}_{\chi V\chi V}$ & ${\cal F}^{m, \mathbf{r}}_{\xi V^\dag\xi V^\dag}$ & ${\cal F}^{m, \mathbf{r}}_{\chi V^\dag\chi V^\dag}$ \\
\midrule\midrule
$m=s$ & $\mathbf{r}={\tiny{\yng(1)}}$ &  & $1$ & & 1 & & &  \\
\hline
 & $\mathbf{r}=\mathbf{1}$ & & & & & & & $N$ \\
\hline
 & $\mathbf{r}=\mathbf{Adj}$ & & & &  &  & 1 & \\
\hline\hline
$m=t$ & $\mathbf{r}=\mathbf{1}$ &  &  & $\sqrt{N}$ & & & &  \\
\hline
 & $\mathbf{r}=\mathbf{Adj}$ &  &  & 1 & & & &  \\
\hline\hline
$m=u$ & $\mathbf{r}={\tiny{\yng(1)}}$ & 1 &  &  & & & &  \\
\hline
 & $\mathbf{r}=\bar{\tiny{\yng(1)}}$ & & & & & & 1 &  \\
\hline
 & $\mathbf{r}={\rm {\bf{S}}}$ & & & & & 1 & & \\
\hline
 & $\mathbf{r}={\rm {\bf{AS}}}$ & & & & & $-1$ & & \\
\end{tabular}
\end{center}
\caption{Group factors coefficient for the second model of the complex vector case. Here $\chi\sim {\tiny{\fbox{$\protect\phantom s$} }}_q$, 
$\xi\sim \mathbf{1}_{q-q^\prime}$,  $V\sim {\tiny{\fbox{$\protect\phantom s$} }}_{q^\prime}$.}
\label{GF2}
\end{table}

\begin{table}[t!]
\begin{center}
\begin{tabular}{cc || c|c|c| c|c|c|c}
\multicolumn{2}{c||}{{\bf Third model}} & ${\cal F}^{m, \mathbf{r}}_{\chi\xi\chi\xi}$ & ${\cal F}^{m, \mathbf{r}}_{\chi\bar\xi\chi\bar\xi}$ & ${\cal F}^{m, \mathbf{r}}_{\chi\bar\chi\xi\bar\xi}$ & ${\cal F}^{m, \mathbf{r}}_{\xi V\xi V}$ & ${\cal F}^{m, \mathbf{r}}_{\chi V\chi V}$ & ${\cal F}^{m, \mathbf{r}}_{\xi V^\dag\xi V^\dag}$ & ${\cal F}^{m, \mathbf{r}}_{\chi V^\dag\chi V^\dag}$ \\
\midrule\midrule
$m=s$ & $\mathbf{r}=\mathbf{Adj}$ &  & $\frac{1}{2}$ & & & & & \\
\hline
 & $\mathbf{r}={\tiny{\yng(1)}}$ & & & & $\frac{N^2-1}{2N}$ & & & $\frac{N^2-1}{2N}$ \\
\hline\hline
$m=t$ & $\mathbf{r}=\mathbf{1}$ &  &  & $\frac{N^2-1}{2N}$ & & & & \\
\hline
 & $\mathbf{r}=\mathbf{Adj}$ &  &  & $-\frac{1}{2N}$ & & & & \\
\hline\hline
$m=u$ & $\mathbf{r}={\rm {\bf{S}}}$ & $\frac{N-1}{2N}$ &  & & & & &  \\
\hline
 & $\mathbf{r}={\rm {\bf{AS}}}$ & $\frac{N+1}{2N}$ &  &  & & & &  \\
\hline
 & $\mathbf{r}={\tiny{\yng(1)}}$ & & & & & $-\frac{1}{2N}$ & $-\frac{1}{2N}$ &  \\
\hline
 & $\mathbf{r}=\mathbf{r}_{1}$ & & & & & $-\frac{1}{2}$ & $-\frac{1}{2}$ & \\
\hline
 & $\mathbf{r}=\mathbf{r}_{2}$ & & & & & $\frac{1}{2}$ & $\frac{1}{2}$ & \\
\end{tabular}
\end{center}
\caption{Group factors coefficient for the third model of the complex vector case. Here $\chi\sim {\tiny{\fbox{$\protect\phantom s$} }}_q$, $\xi\sim {\tiny{\fbox{$\protect\phantom s$} }}_{q^\prime}$, $V\sim \mathbf{Adj}_{q-q^\prime}$}
\label{GF3}
\end{table}

\begin{table}[t!]
\begin{center}
\begin{tabular}{cc || c|c|c| c|c|c|c}
\multicolumn{2}{c||}{{\bf Fourth model}} & ${\cal F}^{m, \mathbf{r}}_{\chi\xi\chi\xi}$ & ${\cal F}^{m, \mathbf{r}}_{\chi\bar\xi\chi\bar\xi}$ & ${\cal F}^{m, \mathbf{r}}_{\chi\bar\chi\xi\bar\xi}$ & ${\cal F}^{m, \mathbf{r}}_{\xi V\xi V}$ & ${\cal F}^{m, \mathbf{r}}_{\chi V\chi V}$ & ${\cal F}^{m, \mathbf{r}}_{\xi V^\dag\xi V^\dag}$ & ${\cal F}^{m, \mathbf{r}}_{\chi V^\dag\chi V^\dag}$ \\
\midrule\midrule
$m=s$ & $\mathbf{r}={\tiny{\yng(1)}}$ &  & $\frac{N^2-1}{2N}$ & & $\frac{N^2-1}{2N}$ & & &  \\
\hline
 & $\mathbf{r}=\mathbf{Adj}$ & & & & & & & $\frac{1}{2}$ \\
\hline\hline
$m=t$ & $\mathbf{r}=\mathbf{1}$ &  &  & $\sqrt{\frac{N^2-1}{4N}}$ & & & &  \\
\hline
 & $\mathbf{r}=\mathbf{Adj}^f$ &  &  & $i\sqrt{\frac{N}{8}}$ & & & & \\
\hline
 & $\mathbf{r}=\mathbf{Adj}^d$ &  &  & $\sqrt{\frac{N^2-4}{8N}}$ & & & & \\
\hline\hline
$m=u$ & $\mathbf{r}={\tiny{\yng(1)}}$ & $-\frac{1}{2N}$ &  & & & & &  \\
\hline
 & $\mathbf{r}=\mathbf{R}_{1}$ & $-\frac{1}{2}$ &  & & & & & \\
\hline
 & $\mathbf{r}=\mathbf{R}_{2}$ & $\frac{1}{2}$ &  & & & & & \\
\hline
 & $\mathbf{r}=\bar{\tiny{\yng(1)}}$ & & & & & & $-\frac{1}{2N}$ &  \\
\hline
 & $\mathbf{r}=\bar{\mathbf{r}}_{1}$ & & & & & & $-\frac{1}{2}$ & \\
\hline
 & $\mathbf{r}=\bar{\mathbf{r}}_{2}$ & & & & & & $\frac{1}{2}$ & \\
\hline
 & $\mathbf{r}={\rm {\bf{S}}}$ & & & & $1$ &  &   \\
\hline
 & $\mathbf{r}={\rm {\bf{AS}}}$ & & & & $-1$ &  &   \\
\end{tabular}
\end{center}
\caption{Group factors coefficient for the fourth model of the complex vector case. Here $\chi\sim {\tiny{\fbox{$\protect\phantom s$} }}_q$, 
$\xi\sim \mathbf{Adj}_{q-q^\prime}$,  $V\sim {\tiny{\fbox{$\protect\phantom s$} }}_{q^\prime}$}
\label{GF4}
\end{table}

\subsubsection{Third model: $\chi\sim {\tiny{\fbox{$\protect\phantom s$} }}_q$, $\xi \sim {\tiny{\fbox{$\protect\phantom s$} }}_{q^\prime}$, $V\sim \mathbf{Adj}_{q-q^\prime}$ }\label{sec:model3}
In this model with $q\ne q^\prime$ $V$ is a complex vector field. In the half-integer $J$ scattering matrix we have the non trivial decomposition
\begin{gather}
{\tiny{\yng(1)}}\times\mathbf{Adj}={\tiny{\yng(1)}}+\mathbf{r}_{1}+\mathbf{r}_{2} \ ,
\end{gather}
where $\mathbf{r}_{1,2}$ are the two irreducible representations arising from the tensor decomposition, with $\mathbf{r}_{1}$ having dimension $N(N+1)(N-2)/2$ and $\mathbf{r}_{2}$ having dimension $N(N-1)(N+2)/2$. The explicit expression for the tensor decomposition is reported in the Appendix of~\cite{Allwicher:2021rtd}.
The group factors for this model are reported in Tab.~\ref{GF3} and the strongest PU constraint turns out to be
\be
g\lesssim\text{Min}\left\{ 1.65\sqrt{\frac{2N}{N+1}}, \ 1.98\sqrt{\frac{2N}{N^2-1}} \right\} \, .
\ee

\subsubsection{Fourth model: $\chi\sim {\tiny{\fbox{$\protect\phantom s$} }}_q$, $\xi \sim \mathbf{Adj}_{q-q^\prime}$, $V \sim {\tiny{\fbox{$\protect\phantom s$} }}_{q^\prime}$}\label{sec:model4}

As in the model of Sec.~\ref{sec:model2} also in this case the vector is always a complex field. The decomposition of the two particle states into irreducible $SU(N)$ representation is similar to the previous models, while the group factors are reported in Tab.~\ref{GF4}.
The strongest constraint on the coupling $g$ in this model is then
\be
g\lesssim\text{Min}\left\{ \frac{2.8}{\sqrt{N}}, \ \sqrt{8\pi}\sqrt{\frac{2N}{N^2-1}} \right\} \, .
\ee

\section{Vector fields from gauge theories}\label{sec:gauge}

In this Section we compute the constraint arising from the requirement of PU on the gauge couplings of a $SU(N)\otimes U(1)$ gauge theory. Here the vector fields are the $SU(N)$ gauge fields $V_{\mu}^{A=1,...,N^2-1}$ transforming as a real adjoint representation of $SU(N)$, and whose self coupling are fixed by the gauge symmetry itself, and the $U(1)$ gauge field $V_{\mu}$ being a real $SU(N)$ singlet. We also include one fermion field $\chi$ transforming as a $SU(N)$ fundamental representation and with charge $q=1$ under $U(1)$.
The interaction Lagrangian of the theory is then
\begin{align}
{\cal L}_{\text{gauge}}&={\cal L}_{\psi}+{\cal L}_{\text{self}} \, , \nn \\
{\cal L}_{\psi}&=g'\bar{\chi}^{a}\gamma^{\mu}\chi_{a}V_{\mu}+g\bar{\chi}^{a}\gamma^{\mu}(T^{A})_{a}^{.\;b}\chi_{b}V_{\mu}^{A} \, , \nn \\
\label{gaugeself}
{\cal L}_{\text{self}}&=-gf_{ABC}(\partial^{\mu}V^{\nu A})V_\mu^B V_\nu^C-\frac{1}{4}g^2(f^{EAB}V_\mu^A V_\nu^B)(f^{ECD}V^{\mu C}V^{\nu D}) \, ,
\end{align}
where $f^{ABC}$ are the $SU(N)$ structure constants and $(T^{A})_{a}^{.\;b}$ are the generators of the $SU(N)$ group. Here $g$ is the $SU(N)$ gauge coupling while $g'$ is the $U(1)$ gauge coupling.
To simplify our analysis, we evaluate the constraint from the PU requirement on one of the gauge couplings while turning off the other one. In the case that more than one gauge coupling is present, the building blocks we provide are enough to derive combined PU bounds, as explicitly shown in~\cite{Allwicher:2021rtd} for the Yukawa coupling case.

\subsection{PU bound on the $SU(N)$ gauge coupling}

In this Section we present the results for the case of gauge fields of an $SU(N)$ group, again dividing  between integer and half-integer $J$ scatterings.

\subsubsection{Integer $J$ scattering: $\{\psi\psi,VV  \}\to \{\psi \psi,VV\}$}

These scattering involve initial and final states with two fermions or two vectors. The scattering matrices in the helicity form are
\vskip 10pt
\be\label{eq:all}
    {\cal T}_{\{\psi\psi,VV\}} =
     \left(
    \begin{array}{c c c !{\vrule width 1pt} c c | c | c c|c}
        &   &  &   \multicolumn{3}{c|}{\psi\psi} &   \multicolumn{3}{c}{VV} \\

         &  &  &   \multicolumn{2}{c|}{\mu_i=0} &  \mu_i=-1   &   \multicolumn{2}{c|}{\mu_i=0} &  \mu_i=-2   \\
       &  & &  ++ 				& -- & -+ &  ++ 				& -- & -+  \\
\specialrule{1pt}{0pt}{0pt}
\multicolumn{1}{c}{}  & \multirow{2}{*}{$\mu_f=0$}  & ++  & {\cal T}^{++++}  &   &   & & &   \\
\multicolumn{1}{c}{\psi\psi} & &--    &  	 &  {\bf {\cal T}}^{----}  &   & & &   \\
\cline{2-9}
\multicolumn{1}{c}{} & \mu_f=-1 &-+    &  	 &  &  {\bf {\cal T}}^{-+-+}   &  {\cal T}^{-+++}	 & {\cal T}^{-+--} &  {\bf {\cal T}}^{-+-+}   \\
\hline
\multicolumn{1}{c}{}  & \multirow{2}{*}{$\mu_f=0$}  & ++  &   &   &   {\cal T}^{++-+}  & {\cal T}^{++++}  &  {\cal T}^{++--} &   {\cal T}^{++-+}   \\
\multicolumn{1}{c}{VV} & &--  &  	 &    &   {\cal T}^{---+}  &  {\cal T}^{--++}	 &  {\bf {\cal T}}^{----}  &   {\cal T}^{---+}   \\
\cline{2-9}
\multicolumn{1}{c}{} & \mu_f=-2 &-+  &  	 &  &  {\bf {\cal T}}^{-+-+}  &  {\cal T}^{-+++}	 & {\cal T}^{-+--} &  {\bf {\cal T}}^{-+-+}      \\
    \end{array}
    \right) \ ,
\ee
together with the one in Eq.~\eqref{eq:full_matrix}. The Lorentz parts for the non zero scatterings are again reported in App.~\ref{sec:LorPart}. Recall that the two particle states with two vectors are kinematically allowed only for $x<1/4$.
Following again Eq.~\eqref{eq:total_T}, we study the partial scattering matrices in a given irreducible representations $\mathbf{r}$ of the group symmetry. The irreducible decomposition of the possible initial or final states are
\begin{gather}
\chi\chi\sim{\rm {\bf{S}}}+{\rm {\bf{AS}}} \, , \nn\\
\chi\bar\chi\sim\mathbf{1}+\mathbf{Adj} \, , \nn\\
VV \sim\mathbf{1}+\mathbf{Adj}^f+\mathbf{Adj}^d+\dots \, ,
\end{gather}
where $f$ and $d$ refer to the two possible contraction, totally antisymmetric and totally symmetric respectively, among two fields in the adjoint.
For $\mathbf{r}={\rm {\bf{S}}},{\rm {\bf{AS}}}$, only the $\chi\chi\to\chi\chi$ scattering is possible and its partial amplitude is given by
\be
a_{\chi\chi\chi\chi}^{J,{\rm {\bf{r}}}}=g^2 f^{(u)}_{J}(x)\left[ {\cal F}^{u, {\rm {\bf{r}}}}_{\chi\chi\chi\chi} + (-1)^J {\cal F}^{t, {\rm {\bf{r}}}}_{\chi\chi\chi\chi} \right] \, ,
\ee
where, by direct computation,
\be
{\cal F}^{t, {\rm {\bf{S}}}}_{\chi\chi\chi\chi}={\cal F}^{u, {\rm {\bf{S}}}}_{\chi\chi\chi\chi}=\frac{N-1}{4N} \, ,\quad \, {\cal F}^{t, {\rm {\bf{AS}}}}_{\chi\chi\chi\chi}=-{\cal F}^{u, {\rm {\bf{AS}}}}_{\chi\chi\chi\chi}=-\frac{N+1}{4N} \, .
\ee
Note that $a_{\chi\chi\chi\chi}^{J,{\rm {\bf{AS}}}}$ vanishes for even values of $J$, while $a_{\chi\chi\chi\chi}^{J,{\rm {\bf{S}}}}$ vanishes for odd values of $J$, as expected~\cite{Jacob:1959at}. Taking the maximal value of the amplitude over the parameters $x$, with the choice $x>0.01$, and $J$, we numerically get the constraint
\be
g\lesssim \text{Min}\left\{ 2.33\sqrt{\frac{N}{N-1}}, \ 3.05\sqrt{\frac{N}{N+1}} \right\} \, .
\ee

\begin{figure}
	\centering
	\includegraphics[scale=0.75]{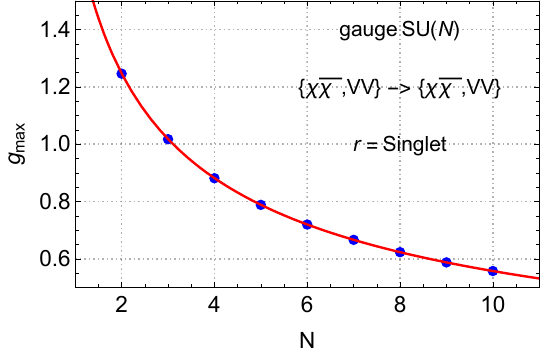}
	\quad
	\includegraphics[scale=0.75]{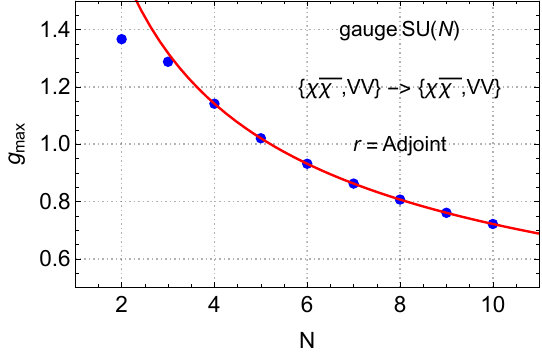}
	\caption{Bound on the $SU(N)$ gauge coupling for the ${\rm {\bf{r}}}=\mathbf{1}$ (left) and ${\rm {\bf{r}}}=\mathbf{Adj}$ (right) as a function of $N$. The red lines represent the asymptotic behaviours of Eq.~\eqref{eq:as_1} and Eq.~\eqref{eq:as_2}.}
	\label{fig:listplotId}
\end{figure}

For $\mathbf{r}=\mathbf{1}$ or $\mathbf{r}=\mathbf{Adj}$, scattering between states with different helicity or particle content are possible and hence the computation is less trivial. Moreover the group decomposition of the states $VV$ contains two independent adjoint representations and both of them communicate with the adjoint representation in the $\chi\bar\chi$ decomposition.
All together, the partial scattering matrices are
\be\label{eq:id}
    a_{\{\chi\bar\chi,VV\}\{\chi\bar\chi,VV\}}^{J,\mathbf{1}} =
     \left(
    \begin{array}{c|c}
a_{\chi\bar\chi\chi\bar\chi}^{J,\mathbf{1}} & a_{\chi\bar\chi \{VV\}}^{J,\mathbf{1}} \\
\hline
a_{\{VV\}\chi\bar\chi}^{J,\mathbf{1}} & a_{\{VV\}\{VV\}}^{J,\mathbf{1}} \\
    \end{array}
    \right) 
\ee
for $\mathbf{r}=\mathbf{1}$ and
\be\label{eq:adj}
    a_{\{\chi\bar\chi,VV\}\{\chi\bar\chi,VV\}}^{J,\mathbf{Adj}} =
     \left(
    \begin{array}{c| c|c}
a_{\chi\bar\chi\chi\bar\chi}^{J,\mathbf{Adj}} & a_{\chi\bar\chi \{VV\}}^{J,\mathbf{Adj}^f} & a_{\chi\bar\chi \{VV\} }^{J,\mathbf{Adj}^d} \\
\hline
a_{ \{VV\} \chi\bar\chi}^{J,\mathbf{Adj}^f} & a_{\{VV\}\{VV\}}^{J,\mathbf{Adj}^f} & 0 \\
\hline
a_{ \{VV\} \chi\bar\chi}^{J,\mathbf{Adj}^d} & 0 & a_{ \{VV\}\{VV\}}^{J,\mathbf{Adj}^d} \\
    \end{array}
    \right) 
\ee
for $\mathbf{r}=\mathbf{Adj}$, where $\{VV\}=\{V_+ V_+,V_-V_-,V_-V_+\}$. The elements of these matrices, together with group factors for $\mathbf{r}=\mathbf{1}$ or $\mathbf{r}=\mathbf{Adj}$, are listed in App.~\ref{App3}. For each irreducible representation, we need to diagonalise the scattering matrix and then the strongest constraint will come from the greatest, in absolute value, eigenvalue. 
We show in the left panel of Fig.~\ref{fig:listplotId} the constraint on the coupling $g$ as a function of $N$, where the
the red line is the asymptotic behaviour at large $N$ estimated as
\be\label{eq:as_1}
g \lesssim \frac{1.76}{\sqrt{N}} \, .
\ee
For ${\rm {\bf{r}}}=\mathbf{Adj}$, we the constraint is shown in the right panel of Fig.~\ref{fig:listplotId} 
where again the red line is the asymptotic behaviour at large $N$ estimated to be
\be\label{eq:as_2}
g \lesssim \frac{2.28}{\sqrt{N}} \, .
\ee

\subsubsection{Half-integer $J$ scattering: $\psi V \to\psi V$}

\begin{figure}[t!]
	\centering
	\includegraphics[scale=0.78]{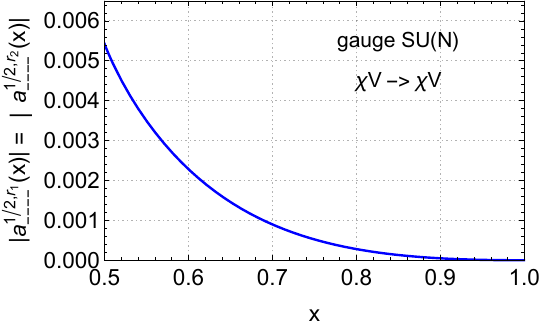}
	\hfill
	\includegraphics[scale=0.78]{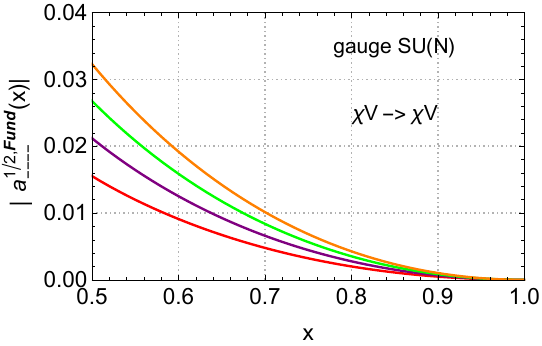}
	\caption{$J=1/2$ partial wave amplitudes for ${\rm {\bf{r}}}=\mathbf{r}_{1,2}$ (left panel) and for the fundamental representation (right panel) with $N=3$ (red), $N=4$ (purple), $N=5$ (green) and $N=6$ (yellow) }
	\label{fig:aaaa}
\end{figure}

For these scatterings, the trilinear self coupling of the gauge vector leads to a $t-$channel contribution. The Lorentz part of the contribution can be found in App~\ref{sec:LorPart}.
By following again Eq.~\eqref{eq:total_T}, the partial scattering matrices in the $\{\chi V_{-1},\chi V_{+1}\}$ basis of in a given representation $\mathbf{r}$ are
\begin{align}
a_{\{\chi V_{-1},\chi V_{+1}\}\{\chi V_{-1},\chi V_{+1}\}}^{J,\mathbf{r}}=&g^2 {\cal F}^{s, {\rm {\bf{r}}}}_{\chi V\chi V}
\begin{pmatrix} b^{(s)}_{J}(x) & 0 \\
0 & 0
\end{pmatrix}
\nn \\
+&g^2 {\cal F}^{u, {\rm {\bf{r}}}}_{\chi V\chi V}
\begin{pmatrix} 
b^{(--)}_{J}(x) & b^{(-+)}_{J}(x)\\
b^{(-+)}_{J}(x) & b^{(++)}_{J}(x)
\end{pmatrix}
\nn \\ 
+&g^2 {\cal F}^{t, {\rm {\bf{r}}}}_{\chi V\chi V}
\begin{pmatrix} 
\beta^{(--)}_{J}(x) & \beta^{(-+)}_{J}(x)\\
\beta^{(-+)}_{J}(x) & \beta^{(++)}_{J}(x)
\end{pmatrix} \, ,
\end{align}
where we have introduced the functions
\begin{align}
& \beta^{(--)}_{J}(x)= -\frac{1}{32\pi}\int_{-1}^{1}\!\text{d}{z}\, P_{J-\frac{1}{2}}^{(0,1)}(z)\frac{(1-x)^{2}(1+z)^2}{2x+(1-x)^2(1-z)} \, , \nn \\
& \beta^{(-+)}_{J}(x)= -\frac{1}{64\pi}\sqrt{\frac{2J+3}{2J-1}}\int_{-1}^{1}\!\text{d}{z}\, P_{J-\frac{3}{2}}^{(2,1)}(z)\frac{(1-x)^{2}(1+z)(1-z)^2}{2x+(1-x)^2(1-z)} \, , \nn \\
& \beta^{(++)}_{J}(x)= -\frac{1}{64\pi}\int_{-1}^{1}\!\text{d}{z}\, P_{J-\frac{3}{2}}^{(0,3)}(z)\frac{(1-x)^{2}(1+z)^3}{2x+(1-x)^2(1-z)} \, .
\end{align}
We apply the same prescription of the complex vector case, thus restricting the analysis to the $J=1/2$ partial amplitudes within the region $1/2<x<1$ to avoid any singularities or divergences. Note that the $\beta^{(-+)}_{J}(x)$ and $\beta^{(++)}_{J}(x)$ functions are absent for $J=1/2$, hence the $2\times2$ matrix is trivially diagonal.
The irreducible decomposition of the initial and final states gives again
\be
{\tiny{\yng(1)}} \otimes \mathbf{Adj}  = {\tiny{\yng(1)}} + \mathbf{r}_{1} + \mathbf{r}_{2} \, ,
\ee
where $\mathbf{r}_{1,2}$ are defined in Sec.~\ref{sec:model3} and~\cite{Allwicher:2021rtd}. For ${\rm {\bf{r}}}=\mathbf{r}_{1,2}$, the group factors are
\be
{\cal F}^{t, \mathbf{r}_{1}}_{\chi V\chi V}=-{\cal F}^{u, \mathbf{r}_{1}}_{\chi V\chi V}=-{\cal F}^{t, \mathbf{r}_{2}}_{\chi V\chi V}={\cal F}^{u, \mathbf{r}_{2}}_{\chi V\chi V}=\frac{1}{2} \, .
\ee
Note that they are independent on $N$. We plot the $J=1/2$ partial amplitudes for ${\rm {\bf{r}}}=\mathbf{r}_{1,2}$ on the panel of Fig.~\ref{fig:aaaa}.
\begin{figure}
	\centering
	\includegraphics[scale=0.76]{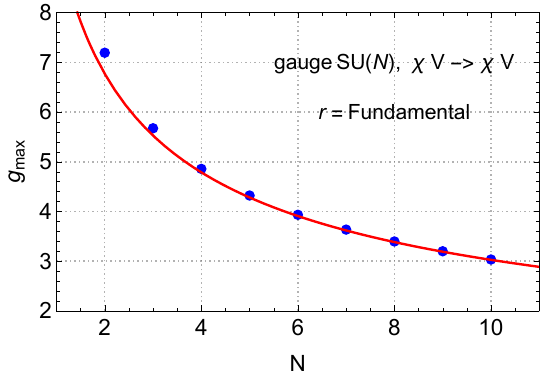}
	\caption{
Bound on the $SU(N)$ gauge coupling for the fundamental representation as a function of $N$. The red lines represent the asymptotic behaviours of Eq.~\eqref{eq:as_1_half_integer}.}
	\label{fig:listplot}
\end{figure}
By taking the maximal value of the amplitude in the allowed region for ${\rm {\bf{r}}}=\mathbf{r}_{1,2}$, we numerically get
\be
g\lesssim 9.6 \, .
\ee
For ${\rm {\bf{r}}}={\tiny{\yng(1)}}$ \ , the group factors are
\be
{\cal F}^{s, {\tiny{\yng(1)}}}_{\chi V\chi V}=\frac{N^2-1}{2N} \, , \, \, {\cal F}^{t, {\tiny{\yng(1)}}}_{\chi V\chi V}=\frac{N}{2} \, , \, \,{\cal F}^{u, {\tiny{\yng(1)}}}_{\chi V\chi V}=-\frac{1}{2N} \, ,
\ee
which depends on $N$.
We plot the $J=1/2$ partial amplitude for ${\rm {\bf{r}}}={\tiny{\yng(1)}}$ on the right panel of Fig.~\ref{fig:aaaa}.
The constraint on the coupling $g$ for ${\rm {\bf{r}}}={\tiny{\yng(1)}}$ depends on $N$ and we show the numerical results in Fig.~\ref{fig:listplot}, where the red line is the asymptotic behaviour at large $N$ estimated as
\be\label{eq:as_1_half_integer}
g \lesssim \frac{9.57}{\sqrt{N}} \, .
\ee

\subsection{PU bound on the $U(1)$ gauge coupling}

We now move to the case of a gauge field of a $U(1)$ group, again dividing  between integer and half-integer $J$ scatterings.

\subsubsection{Integer $J$ scattering: $\{\psi\psi,VV  \}\to \{\psi \psi,VV\}$}

In the case of a scattering in the representations $\mathbf{r}={\rm {\bf{S}}},{\rm {\bf{AS}}}$, only the $\chi\chi\to\chi\chi$ process is possible and its partial amplitude is given by
\be
a_{\chi\chi\chi\chi}^{J,{\rm {\bf{r}}}}=g'^2 f^{(u)}_{J}(x)\left[ {\cal F}^{u, {\rm {\bf{r}}}}_{\chi\chi\chi\chi} + (-1)^J {\cal F}^{t, {\rm {\bf{r}}}}_{\chi\chi\chi\chi} \right] \, ,
\ee
where, by direct computation,
\be
{\cal F}^{t, {\rm {\bf{S}}}}_{\chi\chi\chi\chi}={\cal F}^{u, {\rm {\bf{S}}}}_{\chi\chi\chi\chi}=\frac{1}{2} \, , \, \, {\cal F}^{t, {\rm {\bf{AS}}}}_{\chi\chi\chi\chi}=-{\cal F}^{u, {\rm {\bf{AS}}}}_{\chi\chi\chi\chi}=\frac{1}{2} \, .
\ee
Note that $a_{\chi\chi\chi\chi}^{J,{\rm {\bf{AS}}}}$ vanishes for even values of $J$, while $a_{\chi\chi\chi\chi}^{J,{\rm {\bf{S}}}}$ vanishes for odd values of $J$, as it should be~\cite{Jacob:1959at}. Taking the maximal value of the amplitude over the parameters $x$, again with the choice $x>0.01$, and $J$, we numerically get the constraint
\be
g'\lesssim 1.65 \, .
\ee

\begin{figure}
	\centering
	\includegraphics[scale=0.76]{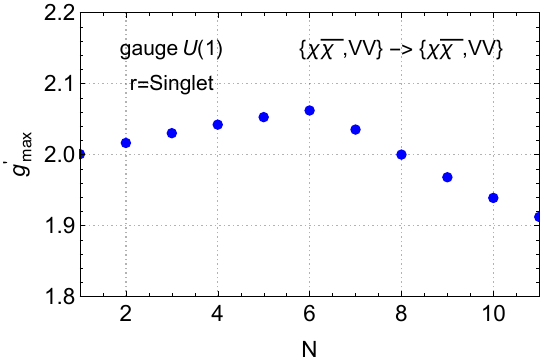}
	\caption{
Bound on the $U(1)$ gauge coupling for the singlet representation as a function of $N$.}
	\label{fig:listplotId_U(1)}
\end{figure}

For $\mathbf{r}=\mathbf{Adj}$, only the $\chi\bar\chi\to\chi\bar\chi$ scattering is possible because $V$ is a singlet and the partial amplitude is given by
\be
a_{\chi\bar\chi\chi\bar\chi}^{J,\mathbf{Adj}}=g'^2 {\cal F}^{t, \mathbf{Adj}}_{\chi\bar\chi\chi\bar\chi} f^{(t)}_{J}(x) \, ,
\ee
with ${\cal F}^{t, \mathbf{Adj}}_{\chi\bar\chi\chi\bar\chi}=1$. Taking the maximal value of the amplitude over the parameters $x$, choosing once again $x>0.01$, and $J$, we numerically get the constraint
\be
g'\lesssim 1.98 \, .
\ee
For $\mathbf{r}=\mathbf{1}$, scattering between states with different helicity or particle content are possible and hence the computation is less trivial. However, due to the absence of a self coupling for the vector field, the partial amplitudes relative to the $VV\to VV$ scattering vanish, hence simplifying the calculation respect the previous case.
For ${\rm {\bf{r}}}=\mathbf{1}$, we plot the constraint on the coupling $g'$ as a function of $N$ in Fig. \ref{fig:listplotId_U(1)}, where at very large $N$ the bound is estimated as
\be
g' \lesssim \frac{6.13}{\sqrt{N}} \, .
\ee

\subsubsection{Half-integer $J$ scattering: $\psi V \to\psi V$}

\begin{figure}[t!]
	\centering
	\includegraphics[scale=0.76]{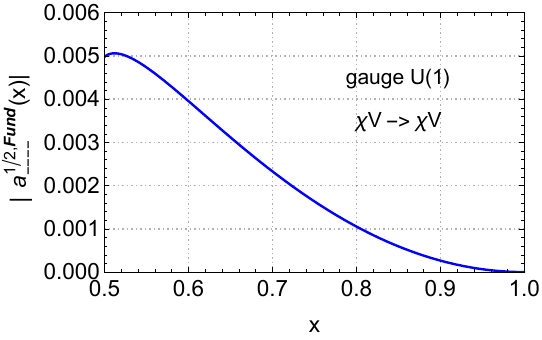}
	\caption{$J=1/2$ partial amplitude for the fundamental representation.}
	\label{fig:aaaaa}
\end{figure}

Following Eq.~\eqref{eq:total_T}, the partial scattering matrices in the $\{\chi V_{-1},\chi V_{+1}\}$ basis in the only possible representation $\mathbf{r}={\tiny{\yng(1)}}$ are
\begin{align}
a_{\{\chi V_{-1},\chi V_{+1}\}\{\chi V_{-1},\chi V_{+1}\}}^{J,{\tiny{\yng(1)}}}=&g'^2 {\cal F}^{s, {\tiny{\yng(1)}}}_{\chi V\chi V}
\begin{pmatrix} b^{(s)}_{J}(x) & 0 \\
0 & 0
\end{pmatrix}
\nn \\
+&g'^2 {\cal F}^{u, {\tiny{\yng(1)}}}_{\chi V\chi V}
\begin{pmatrix} 
b^{(--)}_{J}(x) & b^{(-+)}_{J}(x)\\
b^{(-+)}_{J}(x) & b^{(++)}_{J}(x)
\end{pmatrix} \, ,
\end{align}
where ${\cal F}^{s, {\tiny{\yng(1)}}}_{\chi V\chi V}={\cal F}^{u,{\tiny{\yng(1)}}}_{\chi V\chi V}=1$.
We apply the same prescription of the complex vector case, thus restricting the analysis to the $J=1/2$ partial amplitudes within the region $1/2<x<1$ to avoid any singularities or divergences.
We plot the $J=1/2$ partial amplitude for ${\rm {\bf{r}}}={\tiny{\yng(1)}}$ in Fig.~\ref{fig:aaaaa}.
Taking the maximal value of the amplitude in the allowed region, we numerically get the PU bound
\be
g^\prime \lesssim 9.94 \, .
\ee

\section{Applications}\label{sec:app}

In this Section we apply our results to two illustrative NP models which present an additional vector interaction and which has been formulated to solve anomalies reported in low energy measurements. We focus to the case of the anomalies in charged-current semileptonic decays of neutral $B-$mesons. In particular a tension with the SM prediction is reported for the observables $R{(D^{(*)})}={\rm BR}(B\to D^{(*)}\tau\nu)/{\rm BR}(B\to D^{(*)}\ell\nu)$ with $\ell = e,\mu$, which at the time
after the {\it Winter 2023 update} is around $3.2\sigma$~\cite{HFLAV:2023}. These observables are commonly casted in terms of  the parameters $R_{D^{(*)}}=R{(D^{(*)})}/R{(D^{(*)})}_{\rm SM}$.

\begin{figure}[t!]
	\centering
	\includegraphics[scale=0.78]{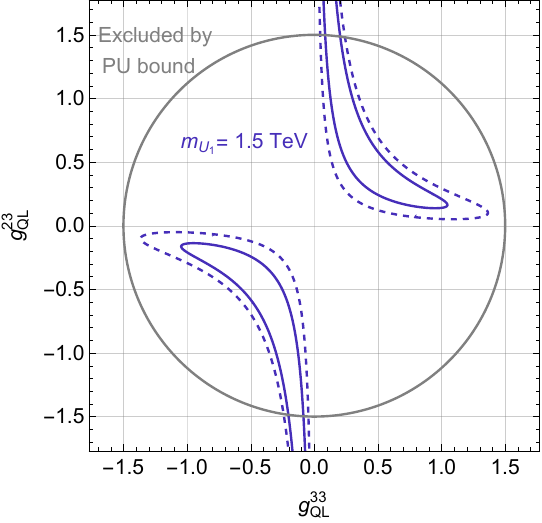}
	\hfill
	\includegraphics[scale=0.78]{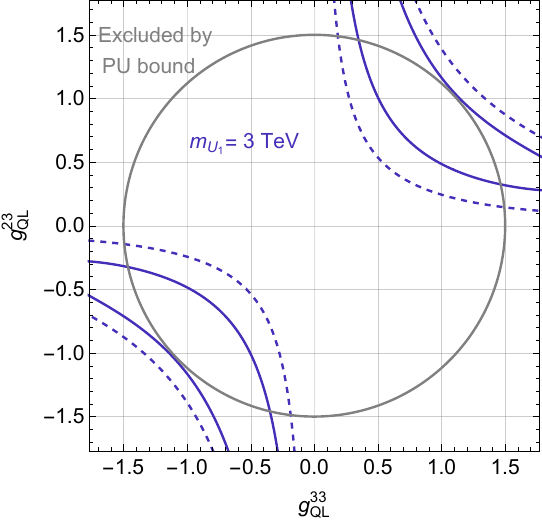}
	\caption{Regions compatible with the $R{(D^{(*)})}$ anomalies for the vector LQ in the model of Eq.~\eqref{eq:lag_U_L} at $1\;\sigma$ (solid) and $2\;\sigma$ (dashed) for $m_{U_1}=1.5\;$TeV (left) and $m_{U_1}=3\;$TeV (right). Outside the gray circle is the region not compatible with the criterium of PU.}
	\label{fig:LQ}
\end{figure}

A common benchmark model is the introduction of a vector leptoquark (LQ) with quantum numbers $({\bf 3},{\bf 1})_{2/3}$ under the SM gauge group coupled to the left-handed SM fermions and known as $U_1^\mu$, see {\emph{e.g.}}~\cite{Buttazzo:2017ixm,DiLuzio:2017vat}.
The effective Lagrangian of the theory is
\be\label{eq:lag_U_L}
{\cal L} = g^{33}_{QL} U_1^\mu \bar Q_L^3  \gamma_\mu L_L^3+g^{23}_{QL} U_1^\mu \bar Q_L^2  \gamma_\mu L_L^3 +h.c.~ \ ,
\ee
with $Q_L^i =(V^*_{ji}u_L^j,d_L^i)$ written in the down-quark basis. Here we assume $g_{QL}^{33}$ and $g_{QL}^{23}$ to be real just for simplicity.
Each interaction in Eq.~\eqref{eq:lag_U_L} correspond to the second type of model of the complex vector case of Sec.~\ref{sec:eff_th} for the $SU(3)$ group and the first type of model of the complex vector case for the $SU(2)$ group.
From Eq.~\eqref{eq:complex} and the group factors of Tab.~\ref{GF1} and Tab.~\ref{GF2} one thus obtains that
\be
\sqrt{(g^{33}_{QL})^2+(g^{23}_{QL})^2}\lesssim1.5 \, .
\ee
The best fit to the charged-current anomaly assuming $R(D)=R(D^*)$ is given by
\be
R_{D^{(*)}} = \left| 1+\frac{v^2g^{33}_{QL}}{2m_U^2 V_{cb}}(g^{33}_{QL}V_{cb}+g^{23}_{QL}V_{cs}) \right|^2 = 1.141 \pm 0.039 \ .
\ee
In Fig.~\ref{fig:LQ} we plot the preferred region of couplings at 1$\sigma$ (continuous line) and $2\sigma$ (dashed line) for different values of the leptoquark mass, also including the constraints from radiative corrections \cite{Buttazzo:2017ixm,Feruglio:2016gvd,Feruglio:2017rjo}. The actual lower limit on the leptoquark mass from high-energy searches is approximately 1.5 TeV \cite{CMS:2020wzx,ATLAS:2023vxj} and for this value of LQ mass there is a large overlap between the preferred region in the $g_{QL}^{33}-g_{QL}^{23}$ parameter space able to explain the $R{(D^{(*)})}$ anomaly and the one compatible with the requirement of PU unitarity, which is inside the gray circle, as illustrated in the left panel. Should direct searches not find any evidence of the $U_1^\mu$ vector LQ and increase the bound on its mass, the models parameters values for which one is able to explain the charged-current $B-$meson anomalies will be pushed at the edge of perturbativity, as shown in the right panel.

Alternatively, one of the Authors proposed in~\cite{Azatov:2018kzb} a NP model that addresses the charged-current anomaly\footnote{At the time the model was able to simultaneously address also the anomaly in neutral-current semileptonic $B$-decays. This anomaly disappeared after the latest results from LHCb~\cite{LHCb:2022qnv,LHCb:2022zom}, showing agreement with the SM predictions.} by adding a new decay channel $B\to D^{(*)}\tau N_R$ into a right-handed sterile neutrino $N_R$, through the exchange of the LQ $U_1$. The Lagrangian of the theory is
\be\label{eq:lag_U}
	{\cal L} = U_1^\mu (g_{cN} \bar c_R  \gamma_\mu N_R +g_{b\tau} \bar b_R \gamma_\mu \tau_R) +h.c.~ \ .
\ee
Each interaction in Eq.~\eqref{eq:lag_U} correspond to the second type of model of the complex vector case of Sec.~\ref{sec:eff_th} given that the RH neutrino is a total SM singlet. For the integer $J$ case the $s-$channel and $u-$channel scatterings occur in the fundamental $SU(3)$ representation, while the $t-$channel one in the singlet one.
From Eq.~\eqref{eq:complex} and the group factors of Tab.~\ref{GF2} one thus obtains that
\begin{align}\label{eq:bound_NR}
&g_{cN,b\tau} \lesssim 1.5 \, ,
\end{align}
while the half-integer $J$ channel enforces a weaker limit. The model must fit the charged-current anomaly that, assuming $R(D)=R(D^*)$ is given by
\be
R_{D^{(*)}} =1+ \frac{|g_{cN}^* g_{b\tau}|^2}{m_{U_1}^4} \frac{v^4}{4 |V_{cb}|^2} = 1.141 \pm 0.039 \ ,
\ee
which implies
\be
|g_{cN}^* g_{b\tau}| \sim 0.50 \left(\frac{m_{U_1}}{1\;{\rm TeV}} \right)^2 \ .
\ee
Given the PU limit of Eq.~\eqref{eq:bound_NR}, this translates into an {\it upper bound} on the mass of the $U_1$ vector LQ which reads
\be\label{eq:bound_U1}
m_{U_1} \lesssim 2.1\;{\rm TeV} \ ,
\ee
while current limits from direct searches requires $m_{U_1} \gtrsim 1.5\;{\rm TeV}$~\cite{Azatov:2018kzb}.
Should the charged current anomaly be confirmed with further measurements, this results implies either that the discovery of the mediator responsible for this observation will be within the reach of the next generation of collider experiments, or that new phenomena beyond perturbativity should be observed at a scale not far from the one indicated by Eq.~\eqref{eq:bound_U1}. In either case this guarantees the discovery of NP with high-energy experiment operating at a center of mass energy above the one of the LHC.

\section{Conclusions}\label{sec:conc}

In this paper we have studied the constraints imposed
by PU on generic vector interactions where the fields involved transform under a $\prod_i SU(N_i) \otimes U(1)$ group. We provide explicit calculation for a simple group $SU(N)$ that can be used as building block to generalise to larger groups such as $\prod_i SU(N_i) $, as shown in Sec. 5 of~\cite{Allwicher:2021rtd}.
By considering all $2\to2$ tree-level scatterings in the high-energy limit we have constructed the general form of the partial-wave matrices $a_{fi}^J$ and derived compact expressions
for the upper limit on the value of the vector interaction up to which perturbation theory
can be trusted.
Differently from the case of Yukawa interactions studied in~\cite{Allwicher:2021rtd}, Coulomb singularities in the tree-level expressions of scattering amplitudes require to consider the vector fields as massive. The PU bound are then a function of the vector mass, 
and can be casted in terms of  the ratio between the mass and scattering energy. We have evaluated the strongest constraint by maximising the partial wave-amplitude over the values of vector mass. For convenience, we  summerize all the results in Tab~\ref{summary}.

We have started by considering a set of phenomenologically relevant toy model involving a complex vector field charged under a $U(1)$ symmetry and where the various fields are only charged under a {\emph{single}} $SU(N)$ factor, working for concreteness in the case where all the fields transform in the trivial, fundamental or adjoint representation. In this class of models, the self coupling among the vector field is neglected and we 
have obtained a bound on the coupling of the vector state to the fermion currents in function of the fields quantum numbers for the various partial wave $J$. We have subsequently considered vector fields from a $SU(N)\otimes U(1)$ gauge theory where the self coupling among the vectors is completely fixed by the gauge symmetry. In order to simplify the analysis we have explicitly evaluated the PU bound on the $SU(N)$ and $U(1)$ gauge couplings separately, by turning on only one of them at time. We have then
applied our results to a phenomenological motivated NP models which adds a vector LQ and a RH neutrino to the SM particle content in order to solve some long standing anomalies reported in semileptonic charged-current decays of $B$ meson. We have in particularly shown that the bounds imposed by perturbative unitarity on the coupling of the theory, together with their best fit to the experimental results, implies either that the discovery of the vector LQ mediator should be possible with the next generation of collider experiments, or that new non perturbative phenomena should appear. We stress that the results presented in this paper are of practical use, and their applicability lies beyond the simple examples presented in the text. While we have restricted our phenomenological analysis to two motivated NP models, 
 the expressions that we have derived furnish the necessary ingredients to study the limits imposed by the requirement of PU in a large set of phenomenologically relevant NP theories that present additional vector interactions.

\begin{table}[t!]
\begin{center}
\begin{tabular}{cc | ccc}
\multicolumn{2}{c|}{{\bf Complex vector}} & \multicolumn{3}{c}{Perturbative unitarity constraint} \\
\midrule
\multicolumn{2}{c|}{First model} &  \multicolumn{3}{c}{$g\lesssim\text{Min}\left\{ 1.65, \
\sqrt{\frac{12\pi}{N}}
\right\}$} \\
\hline
\multicolumn{2}{c|}{Second model} &  \multicolumn{3}{c}{$g\lesssim\text{Min}\left\{ 1.65, \ \frac{1.98}{N^{1/4}}, \ 
\sqrt{\frac{8\pi}{N}}
\right\}$} \\
\hline
\multicolumn{2}{c|}{Third model} &  \multicolumn{3}{c}{$g\lesssim\text{Min}\left\{ 1.65\sqrt{\frac{2N}{N+1}}, \ 1.98\sqrt{\frac{2N}{N^2-1}} \right\}$} \\
\hline
\multicolumn{2}{c|}{Fourth model} &  \multicolumn{3}{c}{$g\lesssim\text{Min}\left\{ \frac{2.8}{\sqrt{N}}, \ \sqrt{8\pi}\sqrt{\frac{2N}{N^2-1}} \right\}$} \\
\hline
\end{tabular}
\end{center}
\vspace{5mm}
\begin{center}
\begin{tabular}{cc | ccc}
\multicolumn{2}{c|}{{\bf Gauge vector}} & \multicolumn{3}{c}{Perturbative unitarity constraint} \\
\midrule
\multicolumn{2}{c|}{$SU(N)$ group} &  \multicolumn{3}{c}{$g\lesssim\frac{1.76}{\sqrt{N}}$} \\
\hline
\multicolumn{2}{c|}{$U(1)$ group} &  \multicolumn{3}{c}{$g^\prime\lesssim\text{Min}\left\{ 1.65,  \frac{6.13}{\sqrt{N}} \right\}$} \\
\hline
\end{tabular}
\end{center}
\caption{{\it Upper table}: summary of the unitary perturbative constraints for all the models of the complex vector case. {\it Lower table}: summary of 
unitary perturbative constraints for $SU(N)$ and $U(1)$ gauge vector cases. Here the bounds refer to the asymptotic behavior at large $N$.}
\label{summary}
\end{table}

\medskip

\section*{Acknowledgements}
This work was supported in part by MIUR under contract PRIN 2017L5W2PT.


\appendix

\section{Wigner $d-$functions}\label{app:wig}

The small Wigner $d-$functions are defined in the angular momentum basis as
\be
d^J_{m, m^\prime}(\theta) = \langle J m^\prime | e^{-i \theta \hat J_y} | J m \rangle  \ ,
\ee
where $\hat J_y$ is the generator of the rotations around the $y-$axis. The explicit expression of these functions used throughout our analysis are
\begin{align}
& d_{0,0}^{J}=P_{J}^{(0,0)}(\cos\theta) \, , \\
& d_{-1,-1}^{J}=\cos^2\frac{\theta}{2}P_{J-1}^{(0,2)}(\cos\theta) \, , \\
& d_{\frac{1}{2},\frac{1}{2}}^{J}=\cos\frac{\theta}{2}P_{J-\frac{1}{2}}^{(0,1)}(\cos\theta) \, , \\
& d_{\frac{1}{2},-\frac{3}{2}}^{J}=d_{-\frac{3}{2},\frac{1}{2}}^{J}=\sqrt{\frac{J+\frac{3}{2}}{J-\frac{1}{2}}}\sin^2\frac{\theta}{2}\cos\frac{\theta}{2}P_{J-\frac{3}{2}}^{(2,1)}(\cos\theta) \, , \\
& d_{-\frac{3}{2},-\frac{3}{2}}^{J}=\cos^3\frac{\theta}{2}P_{J-\frac{3}{2}}^{(0,3)}(\cos\theta) \, , \\
& d_{-1,0}^{J}=\sqrt{\frac{J+1}{J}}\sin\frac{\theta}{2}\cos\frac{\theta}{2}P_{J-1}^{(1,1)}(\cos\theta) \, , \\
& d_{-1,-2}^{J}=-\sqrt{\frac{J+2}{J-1}}\sin\frac{\theta}{2}\cos^3\frac{\theta}{2}P_{J-2}^{(1,3)}(\cos\theta) \, , \\
& d_{0,-2}^{J}=d_{-2,0}^{J}=\sqrt{\frac{(J+2)(J+1)}{J(J-1)}}\sin^2\frac{\theta}{2}\cos^2\frac{\theta}{2}P_{J-2}^{(2,2)}(\cos\theta) \, , \\
& d_{-2,-2}^{J}=\cos^4\frac{\theta}{2} P_{J-2}^{(0,4)}(\cos\theta) \, ,
\end{align}
with the properties
\be
d^J_{m^\prime, m} = (-1)^{m-m^\prime} d^J_{m, m^\prime} = d^J_{-m, -m^\prime} \ .
\ee

\section{Scattering amplitude in the real vector basis}\label{sec:LorPart}

Assuming as interaction Lagrangian ${\cal L} = \lag_{\psi} + \lag_{\rm self}$ with $\lag_{\psi}$ defined in Eq.~\eqref{eq:gen} and 
\beq\label{L2}
{\cal L}_{\text{self}}=-f^{\alpha\beta\gamma}(\partial^{\mu}V^{\nu \alpha})V_\mu^\beta V_\nu^\gamma-\frac{1}{4}(f^{\sigma\alpha\beta}V_\mu^\alpha V_\nu^\beta)(f^{\sigma\gamma\delta}V^{\mu \gamma}V^{\nu \delta}) \, ,
\eeq
where $f^{\alpha\beta\gamma}$ are real and completely antisymmetric, we report in this appendix the Lorentz parts of the amplitude for all the $2\to2$ scattering channels considered in this work
\be
i_1(\lambda_1, p_1) \; i_2 (\lambda_2, p_2) \to f_1 (\lambda_3, p_3)\; f_2 (\lambda_4, p_4) \ .
\ee
 The helicity index $\pm$ has to be intended to be $\pm 1/2$ and $\pm1$ in the case of fermion and vector field respectively. The spinor fields are defined with the same conventions as in~\cite{Allwicher:2021rtd}, while the real vector field reads
\be
A_\mu = \int \frac{d^3 k}{(2\pi)^2 \sqrt{2 E_k}} \sum_{j={\pm1}} \left[  \epsilon_{\mu,j}(k) a_j(k) e^{-i k x} +
\epsilon_{\mu,j}^*(k) a^\dag_j(k) e^{i k x}
 \right] \ ,
\ee
with the transverse polarizations
\be
\epsilon_\pm(p_1) = \frac{1}{\sqrt 2}(0,\mp1,-i,0)  \ ,
\ee
and the one related to $p_{2,3,4}$ defined via rotations.

\subsection{$\psi\psi\to\psi\psi$ scattering}

From explicit computation the Lorentz parts for the non zero scatterings
\be
\psi_i (\lambda_1, p_1) \; \psi_j(\lambda_2, p_2) \to \psi_k(\lambda_3, p_3)\; \psi_l(\lambda_4, p_4)
\ee
are
\begin{align}\label{eq:lor}
{\cal T}^{t u,++++}_{klij} & = ({\cal T}^{t u,----}_{klij})^* = \frac{2s}{t-M^2} {\cal G}_{\alpha i k}{\cal G}_{\alpha j l} +\frac{2s}{u-M^2} {\cal G}_{\alpha i l}{\cal G}_{\alpha j k} \, , \\ 
{\cal T}^{s t,+-+-}_{klij} & = ({\cal T}^{s t,-+-+}_{klij})^*= \frac{2u}{s-M^2+i\Gamma M} {\cal G}_{\alpha i j}{\cal G}_{\alpha l k}+ \frac{2u}{t-M^2} {\cal G}_{\alpha i k}{\cal G}_{\alpha l j}  \, ,\\
{\cal T}^{s u,-++-}_{klij} & =  ({\cal T}^{s u,+--+}_{klij})^*= \frac{2t}{u-M^2} {\cal G}_{\alpha k j}{\cal G}_{\alpha i l}  + \frac{2t}{s-M^2+i\Gamma M} {\cal G}_{\alpha i j} {\cal G}_{\alpha k l}  \, ,
\end{align}
with
\be
\frac{t}{s}=-\sin^2\frac{\theta}{2} \hspace{1cm} \text{and} \hspace{1cm} \frac{u}{s}=-\cos^2\frac{\theta}{2} \, 
\ee
and where $M$ and $\Gamma$ are the mass and the width of the vector field respectively. We have checked that for $M=0$ these amplitudes reproduce,  {\emph{e.g.}}, the M\o ller and Bhabha scatterings amplitudes in QED.

\subsection{$\psi V\to \psi V$ scattering}

From explicit computation the Lorentz parts for the non zero scatterings
\beq
\psi_{i}(\lambda_1,p_1) \; V_{\alpha}(\lambda_2,p_2) \to \psi_{j}(\lambda_3,p_3) \; V_{\beta}(\lambda_4,p_4)
\eeq
are
\begin{align}
\mathcal{T}_{j\beta i\alpha}^{stu,----}=-\left(\mathcal{T}_{j\beta i\alpha}^{stu,++++}\right)^*&=-2{\cal G}_{\alpha ki}{\cal G}_{\beta jk}\frac{s-M^2}{s}\cos\frac{\theta}{2} + \nn \\
&+2if^{\alpha\beta\gamma}{\cal G}_{\gamma ji}\frac{s-M^2}{t-M^2}\cos^3\frac{\theta}{2} + \nn\\
&+2{\cal G}_{\alpha jk}{\cal G}_{\beta ki}\frac{(s-2M^2)(s-M^2)}{us}\sin^{2}\frac{\theta}{2}\cos\frac{\theta}{2}\, ,\\
\mathcal{T}_{j\beta i\alpha}^{tu,-+-+}=-\left(\mathcal{T}_{j\beta i\alpha}^{tu,+-+-}\right)^*&=2if^{\alpha\beta\gamma}{\cal G}_{\gamma ji}\frac{s-M^2}{t-M^2}\cos^3\frac{\theta}{2} +  \nn\\
&+2{\cal G}_{\alpha jk}{\cal G}_{\beta ki}\frac{(s-M^2)}{u}\cos^{3}\frac{\theta}{2}\, ,\\
\mathcal{T}_{j\beta i\alpha}^{tu,-+--}=-\left(\mathcal{T}_{j\beta i\alpha}^{tu,+-++}\right)^*&=2if^{\alpha\beta\gamma}{\cal G}_{\gamma ji}\frac{s-M^2}{t-M^2}\cos\frac{\theta}{2} \sin^2\frac{\theta}{2}+ \nn \\
&+2{\cal G}_{\alpha jk}{\cal G}_{\beta ki}\frac{M^2(s-M^2)}{us}\sin^{2}\frac{\theta}{2}\cos\frac{\theta}{2}\, ,\\
\mathcal{T}_{j\beta i\alpha}^{tu,---+}=-\left(\mathcal{T}_{j\beta i\alpha}^{tu,+++-}\right)^*&=2if^{\alpha\beta\gamma}{\cal G}_{\gamma ji}\frac{s-M^2}{t-M^2}\cos\frac{\theta}{2} \sin^2\frac{\theta}{2}+ \nn \\
&+2{\cal G}_{\alpha jk}{\cal G}_{\beta ki}\frac{M^2(s-M^2)}{us}\sin^{2}\frac{\theta}{2}\cos\frac{\theta}{2}\, ,
\end{align}
with
\be
\frac{t}{s}=-\left(\frac{s-M^2}{s}\right)^2\sin^2\frac{\theta}{2} \, 
\ee
and
\be
\frac{u}{s}=\frac{M^4}{s^2}-\left(\frac{s-M^2}{s}\right)^2\cos^2\frac{\theta}{2} \, .
\ee

\subsection{$\psi \psi \to V V$ and $VV\to\psi\psi$ scatterings}

From explicit computation the Lorentz parts for the non zero scatterings
\beq
\psi_{i}(\lambda_1,p_1) \; \psi_{j}(\lambda_2,p_2) \to V_{\alpha}(\lambda_3,p_3) \; V_{\beta}(\lambda_4,p_4)
\eeq
and
\beq
V_\alpha(\lambda_1,p_1) \; V_{\beta}(\lambda_2,p_2) \to \psi_{i}(\lambda_3,p_3) \; \psi_{j}(\lambda_4,p_4)
\eeq
are
\begin{align}
\mathcal{T}_{\alpha\beta ij}^{stu,---+}=-\left(\mathcal{T}_{\alpha\beta ij}^{stu,+++-}\right)^*=&\mathcal{T}_{ij\alpha\beta}^{stu,+-++}=-\left(\mathcal{T}_{ij\alpha\beta}^{stu,-+--}\right)^*= \nn\\
=&2{\cal G}_{\gamma ji}if^{\alpha\beta\gamma}\frac{\sqrt{s(s-4M^2)}}{s-M^2}\sin\frac{\theta}{2}\cos\frac{\theta}{2} + \nn\\
-&{\cal G}_{\alpha ki}{\cal G}_{\beta jk}\frac{s}{t}\sin\frac{\theta}{2}\cos\frac{\theta}{2}\left[ 2\sin^{2}\frac{\theta}{2}-1+\sqrt{\frac{s-4M^2}{s}} \right] + \nn\\
+&{\cal G}_{\alpha jk}{\cal G}_{\beta ki}\frac{s}{u}\sin\frac{\theta}{2}\cos\frac{\theta}{2}\left[ 2\cos^{2}\frac{\theta}{2}-1+\sqrt{\frac{s-4M^2}{s}} \right] \, , \\
\mathcal{T}_{\alpha\beta ij}^{stu,++-+}=-\left(\mathcal{T}_{\alpha\beta ij}^{stu,--+-}\right)^*=&\mathcal{T}_{ij\alpha\beta}^{stu,+---}=-\left(\mathcal{T}_{ij\alpha\beta}^{stu,-+++}\right)^*= \nn\\
=&2{\cal G}_{\gamma ji}if^{\alpha\beta\gamma}\frac{\sqrt{s(s-4M^2)}}{s-M^2}\sin\frac{\theta}{2}\cos\frac{\theta}{2} + \nn  \\
+&{\cal G}_{\alpha ki}{\cal G}_{\beta jk}\frac{s}{t}\sin\frac{\theta}{2}\cos\frac{\theta}{2}\left[ 2\cos^{2}\frac{\theta}{2}-1-\sqrt{\frac{s-4M^2}{s}} \right] + \nn \\
-&{\cal G}_{\alpha jk}{\cal G}_{\beta ki}\frac{s}{u}\sin\frac{\theta}{2}\cos\frac{\theta}{2}\left[ 2\sin^{2}\frac{\theta}{2}-1-\sqrt{\frac{s-4M^2}{s}} \right] \, , \\
\mathcal{T}_{\alpha\beta ij}^{tu,-+-+}=-\left(\mathcal{T}_{\alpha\beta ij}^{tu,+-+-}\right)^*=&\mathcal{T}_{ij\alpha\beta}^{tu,+-+-}=-\left(\mathcal{T}_{ij\alpha\beta}^{tu,-+-+}\right)^*= \nn\\
=&-2{\cal G}_{\alpha ki}{\cal G}_{\beta jk}\frac{s}{t}\sin\frac{\theta}{2}\cos^3\frac{\theta}{2}-2{\cal G}_{\alpha jk}{\cal G}_{\beta ki}\frac{s}{u}\sin\frac{\theta}{2}\cos^3\frac{\theta}{2} \, , \\
\mathcal{T}_{\alpha\beta ij}^{tu,+--+}=-\left(\mathcal{T}_{\alpha\beta ij}^{tu,-++-}\right)^*=&\mathcal{T}_{ij\alpha\beta}^{tu,+--+}=-\left(\mathcal{T}_{ij\alpha\beta}^{tu,-++-}\right)^*= \nn\\
=&2{\cal G}_{\alpha ki}{\cal G}_{\beta jk}\frac{s}{t}\sin^3\frac{\theta}{2}\cos\frac{\theta}{2}+2{\cal G}_{\alpha jk}{\cal G}_{\beta ki}\frac{s}{u}\sin^3\frac{\theta}{2}\cos\frac{\theta}{2} \, ,
\end{align}
with
\be
t=M^2+\frac{\sqrt{s(s-4M^2)}-s}{2}-\sqrt{s(s-4M^2)}\sin^2\frac{\theta}{2}
\ee
and
\be
u=M^2+\frac{\sqrt{s(s-4M^2)}-s}{2}-\sqrt{s(s-4M^2)}\cos^2\frac{\theta}{2} \, .
\ee

\subsection{$V V \to V V$ scattering}

From explicit computation the Lorentz parts for the non zero scatterings
\beq
V_\alpha(\lambda_1,p_1) \; V_{\beta}(\lambda_2,p_2) \to V_\gamma(\lambda_3,p_3) \; V_\delta(\lambda_4,p_4)
\eeq
are
\begin{align}
&\mathcal{T}_{\gamma\delta\alpha\beta}^{stu,\lambda_3 \lambda_4 \lambda_1 \lambda_2}= \nn\\
=&f^{\alpha\beta\sigma}f^{\gamma\delta\sigma}\left[ \frac{u-t}{s-M^2}\frac{1+\lambda_1 \lambda_2}{2}\frac{1+\lambda_3 \lambda_4}{2} + \frac{1-\lambda_1 \lambda_4 \cos\theta}{2}\frac{1-\lambda_2 \lambda_3 \cos\theta}{2} - \frac{1+\lambda_1 \lambda_3 \cos\theta}{2}\frac{1+\lambda_2 \lambda_4 \cos\theta}{2}\right] + \nn\\
+&f^{\alpha\gamma\sigma}f^{\delta\beta\sigma}\left[ \frac{s-u}{t-M^2}\frac{1+\lambda_1 \lambda_3 \cos\theta}{2}\frac{1+\lambda_2 \lambda_4 \cos\theta}{2}+\frac{1+\lambda_1 \lambda_2}{2}\frac{1+\lambda_3 \lambda_4}{2}-\frac{1-\lambda_1 \lambda_4 \cos\theta}{2}\frac{1-\lambda_2 \lambda_3 \cos\theta}{2}\right] + \nn\\
+&f^{\alpha\delta\sigma}f^{\beta\gamma\sigma}\left[ \frac{t-s}{u-M^2}\frac{1-\lambda_1 \lambda_4 \cos\theta}{2}\frac{1-\lambda_2 \lambda_3 \cos\theta}{2}+\frac{1+\lambda_1 \lambda_3 \cos\theta}{2}\frac{1+\lambda_2 \lambda_4 \cos\theta}{2}-\frac{1+\lambda_1 \lambda_2}{2}\frac{1+\lambda_3 \lambda_4}{2}\right] \, ,
\end{align}
with
\be
t=-(s-4M^2)\sin^2\frac{\theta}{2}
\ee
and
\be
u=-(s-4M^2)\cos^2\frac{\theta}{2} \, .
\ee


\section{Functions and group factors for the gauge vectors}\label{App3}

In this appendix we report the group factors and the relevant functions entering the computation of the PU bounds for the gauge field cases of Sec.~\ref{sec:gauge}. 

\begin{align}
& a_{\chi\bar\chi\chi\bar\chi}^{J,{\rm {\bf{r}}}}= g^2 {\cal F}^{s, {\rm {\bf{r}}}}_{\chi\bar\chi\chi\bar\chi} f^{(s)}_{J}(x,\gamma) + g^2 {\cal F}^{t, {\rm {\bf{r}}}}_{\chi\bar\chi\chi\bar\chi} f^{(t)}_{J}(x) \, , \\
& a_{V_\mp V_\mp\chi\bar\chi}^{J,{\rm {\bf{r}}}}=g^2 {\cal F}^{s, {\rm {\bf{r}}}}_{VV\chi\bar\chi} h^{(s)}_{J}(x,\gamma) + g^2 \left[ {\cal F}^{t, {\rm {\bf{r}}}}_{VV\chi\bar\chi} - (-1)^J {\cal F}^{u, {\rm {\bf{r}}}}_{VV\chi\bar\chi} \right] h^{(tu)}_{J}(x) \, , \\
& a_{\chi\bar\chi V_\mp V_\mp}^{J,{\rm {\bf{r}}}}=\left( a_{V_\mp V_\mp\chi\bar\chi}^{J,{\rm {\bf{r}}}} \right)^* \, , \\
& a_{V_- V_+\chi\bar\chi}^{J,{\rm {\bf{r}}}}=g^2 {\cal F}^{t, {\rm {\bf{r}}}}_{VV\chi\bar\chi} h^{(t)}_{J}(x) + g^2 {\cal F}^{u, {\rm {\bf{r}}}}_{VV\chi\bar\chi} h^{(u)}_{J}(x) \, , \\
& a_{\chi\bar\chi V_- V_+}^{J,{\rm {\bf{r}}}}=\left( a_{V_- V_+\chi\bar\chi}^{J,{\rm {\bf{r}}}} \right)^* \, , \\
& a_{V_{\lambda_{3}}V_{\lambda_{4}}V_{\lambda_{1}}V_{\lambda_{2}}}^{J,{\rm {\bf{r}}}}= g^2 \sqrt{1-4x}\sum_{m=s,t,u} {\cal F}^{m, {\rm {\bf{r}}}}_{V_{\lambda_{3}}V_{\lambda_{4}}V_{\lambda_{1}}V_{\lambda_{2}}} v^{(\lambda_{3}\lambda_{4})(\lambda_{1}\lambda_{2})}_{m,J}(x) \, ,
\end{align}
where
\begin{align}
& h^{(s)}_{J}(x,\gamma)=\frac{\sqrt{2}}{48\pi}\frac{(1-4x)^{\frac{3}{4}}}{1-x+i\gamma x}\delta_{J1} \, , \\
& h^{(tu)}_{J}(x)=\frac{(1-4x)^{\frac{1}{4}}}{64\pi}\sqrt{\frac{J+1}{J}}\int_{-1}^{1}\!\text{d}{z}\, P_{J-1}^{(1,1)}(z)\frac{z-\sqrt{1-4x}}{2x-1+z\sqrt{1-4x}} \, , \\
& h^{(t)}_{J}(x)=\frac{(1-4x)^{\frac{1}{4}}}{128\pi}\int_{-1}^{1}\!\text{d}{z}\, P_{J-2}^{(1,3)}(z)\frac{(1-z)(1+z)^3}{2x-1+z\sqrt{1-4x}} \, , \\
 & h^{(u)}_{J}(x) =\frac{(1-4x)^{\frac{1}{4}}}{128\pi}\int_{-1}^{1}\!\text{d}{z}\, P_{J-2}^{(1,3)}(z)\frac{(1-z)(1+z)^3}{2x-1-z\sqrt{1-4x}} \, , \\
& v^{(\lambda_{3}\lambda_{4})(\lambda_{1}\lambda_{2})}_{s,J}(x)=-\delta_{\lambda_1\lambda_2}\delta_{\lambda_3\lambda_4}\delta_{J1}\frac{1}{48\pi}\left[ \lambda_1\lambda_3+\frac{1-4x}{1-x} \right] \, , \\
& v^{(\lambda_{3}\lambda_{4})(\lambda_{1}\lambda_{2})}_{t,J}(x)=\delta_{\lambda_1\lambda_2}\delta_{\lambda_3\lambda_4}\frac{1}{48\pi}\left[ 3\delta_{J0}+\lambda_1\lambda_3\delta_{J1} \right]  + \nn \\
& -\frac{1}{16\pi}\int_{-1}^{1}\!\text{d}{z}\,d_{\lambda_1-\lambda_2,\lambda_3-\lambda_4}^{J}(\cos^{-1}z) \frac{1+\lambda_1 \lambda_3 z}{2}\frac{1+\lambda_2 \lambda_4 z}{2}\frac{2-3x}{1-2x-z(1-4x)} \, , \\
& v^{(\lambda_{3}\lambda_{4})(\lambda_{1}\lambda_{2})}_{u,J}(x)=\delta_{\lambda_1\lambda_2}\delta_{\lambda_3\lambda_4}\frac{1}{48\pi}\left[ -3\delta_{J0}+\lambda_1\lambda_3\delta_{J1} \right] + \nn \\
& +\frac{1}{16\pi}\int_{-1}^{1}\!\text{d}{z}\,d_{\lambda_1-\lambda_2,\lambda_3-\lambda_4}^{J}(\cos^{-1}z) \frac{1-\lambda_1 \lambda_4 z}{2}\frac{1-\lambda_2 \lambda_3 z}{2}\frac{2-3x}{1-2x+z(1-4x)} \, .
\end{align}

The needed non-vanishing group factor are
\begin{align}
& {\cal F}^{t, {\rm {\bf{S}}}}_{\chi\chi\chi\chi}={\cal F}^{u, {\rm {\bf{S}}}}_{\chi\chi\chi\chi}=\frac{N-1}{4N} \, , \, \, {\cal F}^{t, {\rm {\bf{AS}}}}_{\chi\chi\chi\chi}=-{\cal F}^{u, {\rm {\bf{AS}}}}_{\chi\chi\chi\chi}=-\frac{N+1}{4N} \, , \\
& {\cal F}^{t, \mathbf{1}}_{\chi\bar\chi\chi\bar\chi}=\frac{N^2-1}{2N} \, , \, \, {\cal F}^{t, \mathbf{1}}_{V_{\lambda_{3}}V_{\lambda_{4}}\chi\bar\chi}={\cal F}^{u, \mathbf{1}}_{V_{\lambda_{3}}V_{\lambda_{4}}\chi\bar\chi}=\left(\frac{1}{\sqrt{2}}\right)^{\delta_{\lambda_3\lambda_4}}\frac{1}{2}\sqrt{\frac{N^2-1}{N}} \, ,\\ 
& {\cal F}^{t, \mathbf{1}}_{V_{\lambda_{3}}V_{\lambda_{4}}V_{\lambda_{1}}V_{\lambda_{2}}}=-{\cal F}^{u, \mathbf{1}}_{V_{\lambda_{3}}V_{\lambda_{4}}V_{\lambda_{1}}V_{\lambda_{2}}}=-\left(\frac{1}{\sqrt{2}}\right)^{\delta_{\lambda_1\lambda_2}+\delta_{\lambda_3\lambda_4}}N \, , \\
& {\cal F}^{s, \mathbf{Adj}}_{\chi\bar\chi\chi\bar\chi}=\frac{1}{2} \, , \, \, {\cal F}^{t, \mathbf{Adj}}_{\chi\bar\chi\chi\bar\chi}=-\frac{1}{2N} \, , \\
& {\cal F}^{s, \mathbf{Adj}^f}_{V_{\lambda_{3}}V_{\lambda_{4}}\chi\bar\chi}=\left(\frac{1}{\sqrt{2}}\right)^{\delta_{\lambda_3\lambda_4}}i\sqrt{\frac{N}{2}} \, , \, \, {\cal F}^{t, \mathbf{Adj}^f}_{V_{\lambda_{3}}V_{\lambda_{4}}\chi\bar\chi}=-{\cal F}^{u, \mathbf{Adj}^f}_{V_{\lambda_{3}}V_{\lambda_{4}}\chi\bar\chi}=\left(\frac{1}{\sqrt{2}}\right)^{\delta_{\lambda_3\lambda_4}}\frac{i}{2}\sqrt{\frac{N}{2}} \, , \\
& {\cal F}^{s, \mathbf{Adj}^f}_{V_{\lambda_{3}}V_{\lambda_{4}}V_{\lambda_{1}}V_{\lambda_{2}}}=\left(\frac{1}{\sqrt{2}}\right)^{\delta_{\lambda_1\lambda_2}+\delta_{\lambda_3\lambda_4}}N \, , \, \, {\cal F}^{t, \mathbf{Adj}^f}_{V_{\lambda_{3}}V_{\lambda_{4}}V_{\lambda_{1}}V_{\lambda_{2}}}={\cal F}^{u, \mathbf{Adj}^f}_{V_{\lambda_{3}}V_{\lambda_{4}}V_{\lambda_{1}}V_{\lambda_{2}}}=-\left(\frac{1}{\sqrt{2}}\right)^{\delta_{\lambda_1\lambda_2}+\delta_{\lambda_3\lambda_4}}\frac{N}{2} \, , \\
& {\cal F}^{t, \mathbf{Adj}^d}_{V_{\lambda_{3}}V_{\lambda_{4}}\chi\bar\chi}={\cal F}^{u, \mathbf{Adj}^d}_{V_{\lambda_{3}}V_{\lambda_{4}}\chi\bar\chi}=\left(\frac{1}{\sqrt{2}}\right)^{\delta_{\lambda_3\lambda_4}}\frac{1}{2}\sqrt{\frac{N^2-4}{2N}} \, , \\
& {\cal F}^{t, \mathbf{Adj}^d}_{V_{\lambda_{3}}V_{\lambda_{4}}V_{\lambda_{1}}V_{\lambda_{2}}}=-{\cal F}^{u, \mathbf{Adj}^d}_{V_{\lambda_{3}}V_{\lambda_{4}}V_{\lambda_{1}}V_{\lambda_{2}}}=\left(\frac{1}{\sqrt{2}}\right)^{\delta_{\lambda_1\lambda_2}+\delta_{\lambda_3\lambda_4}}\frac{N}{2}\frac{N^2+1}{N^2-1} \, .
\end{align}


\newpage
\bibliographystyle{JHEP}
{\footnotesize
\bibliography{biblibimbi}}
\end{document}